\documentclass[10pt, twocolumn,twoside,pre]{revtex4-1}
\usepackage{color}
\usepackage{siunitx}
\usepackage{mathtools}
\usepackage[normalem]{ulem}
\usepackage{color}

\begin{document}
\title{How the growth of ice depends on the fluid dynamics underneath}

\author{Ziqi Wang}
\affiliation{Center for Combustion Energy, Key Laboratory for Thermal Science and Power Engineering of Ministry of Education, Department of Energy and Power Engineering, Tsinghua University, 100084 Beijing, China}
\author{Enrico Calzavarini} 
\affiliation{Univ.\ Lille, Unit\'e de M\'ecanique de Lille - J. Boussinesq - UML - ULR 7512, F-59000 Lille, France}
\author{Chao Sun}
\affiliation{Center for Combustion Energy, Key Laboratory for Thermal Science and Power Engineering of Ministry of Education, Department of Energy and Power Engineering, Tsinghua University, 100084 Beijing, China}
\affiliation{Department of Engineering Mechanics, School of Aerospace Engineering, Tsinghua University, Beijing 100084, China}
\email{chaosun@tsinghua.edu.cn}
\author{Federico Toschi}
\affiliation{Department of Physics and Department of Mathematics and Computer Science and J. M. Burgers Centre for Fluid Dynamics, Eindhoven University of Technology, 5600 MB Eindhoven, The Netherlands; CNR-IAC, Via dei Taurini 19, 00185 Rome, Italy}

\begin{abstract}
{Convective flows coupled with solidification or melting in water bodies play a major role in shaping geophysical landscapes. Particularly in relation to the global climate warming scenario, it is essential to be able to accurately quantify how water-body environments dynamically interplay with ice formation or melting process. Previous studies have revealed the complex nature of the icing process, but have often ignored one of the most remarkable particularity of water, its density anomaly, and the induced stratification layers interacting and coupling in a complex way in presence of turbulence and phase change. By combining experiments, numerical simulations, and theoretical modeling, we investigate solidification of freshwater, properly considering phase transition, water density anomaly, and real physical properties of ice and water phases, which we show to be essential for correctly predicting the different qualitative and quantitative behaviors. We identify, with increasing thermal driving, four distinct flow-dynamics regimes, where different levels of coupling among ice front, stably and unstably stratified water layers occur. Despite the complex interaction between the ice front and fluid motions, remarkably, the average ice thickness and growth rate can be well captured with the theoretical model. It is revealed that the thermal driving has major effects on the temporal evolution of the global icing process, which can vary from a few days to a few hours in the current parameter regime. Our model can be applied to general situations where the icing dynamics occurs under different thermal and geometrical conditions (e.g. cooling conditions or water layer depth).
}
\end{abstract}



\maketitle

Many geophysical patterns result from the interaction between fluid motions and the dynamical evolution of solid phase boundaries. Usually, the dynamics of the solid boundaries are due to phase change or erosion. Examples range from sculpturing of the glacier, ice shelf, iceberg, and sea caves due to flows in the oceans, to congelation ice forming in ponds and lakes and many geological patterns \cite{meakin2010geological}, astrophysical landforms \cite{Alboussiere2010Melting}, as well as in our daily lives and many industrial processes \cite{epstein1983complex,worster1997convection}.

Generally, warm water (freshwater or water with low enough salinity) is lighter and so it floats, whereas cold water is denser and therefore it sinks. However, this is not the case once water is around the density-peak temperature, $T_\text{c}$ (around 4$^\circ$C), when its density reaches the maximum: water expands when it is colder than $T_\text{c}$ (the nonmonotonic relationship of density with temperature for water near $T_\text{c}$ is reported in {\color{blue} \textit{SI Appendix}, section C and Fig. S3}). During cold weather conditions when the lake is close to freezing, colder water (less than $T_\text{c}$) floats to the top and warmer water (more than $T_\text{c}$) sinks. Consequently, the coldest water, which sits on top of the lake releases heat under cold weather conditions, freezes to form a layer of ice. That is why ice first forms on top of water bodies.
The temperature structure in shallow ice-covered lakes is characterized by a continuous increase from 0$^\circ$C at the ice-water interface up to $T_\text{c}$ or higher at the bottom layers in the deep parts of the lake \cite{bilello1968water}.  A research report on the ice-covered Karelian lake found that at the ice formation, a weak stable stratification existed in the lakes with average temperatures about $1^\circ$C. When there is a strong stratification, the turbulent mixing tends to be suppressed. While the temperature exceeds $T_\text{c}$ the interior convection develops that is important to the flow dynamics in the water beneath ice \cite{malm1998bottom}. 
This water density anomaly results in a complex coupling between the ice layer, the gravitationally stably stratified layer of fluid ($0<T\le T_\text{c}$) and the unstably stratified layer ($T>T_\text{c}$, with convective instability) \cite{Veronis1963Penetrative,arid2012numerical,large2014penetrative,corcione2015penetrative, ayurzana2016phase,2019Penetrative,2020_coupled,lecoanet2015numerical,toppaladoddi2018penetrative}. The stably stratified layer always exists in the ice-water system, but its strength may be enhanced or depleted under different levels of turbulence.  

Connecting to the complex fluid dynamics in the water, the evolution of ice front and the phase change at the interface show very rich dynamics, which recently has received increasing attention. Rayleigh-B\'enard (RB) convection, a fluid layer confined between a cold top plate and a hot bottom plate \cite{sig94,bod00,ahlers2009heat,lohse2010small,chi12}, is an ideal model system to study the aforementioned coupled dynamics. Various studies have been performed on the flow in the RB system with freezing or melting boundary conditions. The focus has been on the behaviors of global quantities such as the heat flux, the kinetic energy, and the dynamics of the ice-water interface morphology with a melting phase-change boundary in the RB system \cite{Madruga2018Dynamic,Esfahani2018Basal,favier_purseed_duchemin_2019,Satbhai2019Comparison}, {\color{black}pattern selection and instability analysis with a moving solid-water interface \cite{davis1983pattern,dietsche1985influence},} the bistability of the equilibria induced by different initial conditions \cite{vasil_proctor_2011,2020_Bistability}, melting in double diffusive convection \cite{Sugawara2000The,Sugawara2007Visual,mergui2002ice}, the influences of different container shapes on the melting and convection of phase change materials \cite{Dhaidan2015Melting,hu2017lattice, sugawara2008melting}.

While most of these studies consider the interaction of phase change with the convective motions in the fluid, yet several crucial ingredients have not been fully taken into consideration, notably the water density anomaly and the real physical properties of the ice and water. These ingredients are crucial to realistically capture the growth of the ice layer and the dynamical coupling mentioned above. For example, in geophysical flows, with a typical water temperature in winter of the range $0 \sim 15 ^\circ$C (see examples of Historical Lake Erie Temperatures from National weather service \cite{wintertemp}), it is essential to consider the realistic natural configurations to make correct predictions, e.g., how thick the ice can form and how long it takes to arrive at the equilibrium state for a given environmental condition. What's more, correctly predicting the ice formation time scale can provide a natural indicator of climate change \cite{magnuson2000historical,magee2016trends,preston2016climate,magee2017effects}.

In this work, we combine experiments, numerical simulations, and theoretical modeling to study the coupled dynamics of freshwater solidification and the surrounding fluid dynamics, properly accounting for the water density anomaly and the real physical properties of ice and water. We aim to reveal how the growth of freshwater ice depends on the environmental conditions.

\section*{Results and Discussion}
\subsection*{Experiments and simulations}
The experiments are performed in a Rayleigh-B\'enard convection system of cuboid shape (aspect ratio $\Gamma = L_x / H = 1$, $L_x$, H are the system width and height)
heated up from the bottom and cooled down from the top. Water, as the working fluid, is deionized, ultrapured and degassed. The top plate temperature, $T_\text{t}$, and bottom plate temperature, $T_\text{b}$, are imposed by water-circulating bath, with $T_\text{t}<T_\phi$ and $T_\text{b} > T_\phi$ ($T_\phi$ is the water freezing point, $T_\phi=0^\circ$C). In such a configuration, ice starts forming from the top plate and it grows till its saturation thickness. During the experimental process, there is a volume change induced by thermal expansion of water and water-ice phase change, so an open expansion vessel is connected to the experimental cell allowing to quantify the volume change, and therefore the pressure of the system remains atmospheric pressure. By monitoring the water volume change inside the expansion vessel, the evolution of the spatial average ice thickness can also be calculated (details are shown in {\color{blue}\textit{SI Appendix}, section A and B and Figs. S1 and Figs. S2}). In addition to the experiments, the numerical simulations are carried out using Lattice-Boltzmann method (LBM) numerical code \cite{succi2001lattice,huber2008lattice,Esfahani2018Basal}. In the simulations, we consider the density anomaly, the source term from the latent heat at the ice front \cite{Moritz2019An}, and the correction for the governing equations when the investigated domain consists of heterogeneous media, i.e., ice and water phases ({\color{blue} \textit{SI Appendix}, section E}) \cite{chen2017a}. Two- ($\Gamma = L_x/H = 1$) and three-dimensional simulations are conducted ($\Gamma = L_x/H = 1, L_y = H/4$, same as the experimental cell; $L_y$ being the system width), and the boundary conditions are no-slip for the velocity, adiabatic at the sidewalls, and constant temperatures at the top and bottom plates. The initial condition is still fluid at uniform temperature, $T_\text{b}$. We assume thermophysical properties to be constant except for the density in the buoyancy term. The real water density property near to $T_\text{c}$ is well described with the equation $ \rho=\rho_0(1-\alpha^*|T-T_\text{c}|^q )$, where $\alpha^*$ is not the usual thermal expansion coefficient but has units of $K^{-q}$ with $q=1.895$ and $\alpha^* = 9.30\times10^{-6} (K^{-q})$. This equation gives the maximum density of water $\rho_0 = 999.972~kg/m^3$ at $T = T_\text{c}$ \cite{Gebhart1977A} (see also in {\color{blue} \textit{SI Appendix}, section C and Fig. S3}).

One important control parameter of the system is the Rayleigh number, Ra$_\text{e}$, which is the dimensionless thermal forcing, and its definition formula is explained below (more details are shown in {\color{blue} \textit{SI Appendix}, section D}). Another important control parameter is the Stefan number which relates the sensible heat to the latent heat, $\text{Ste} = L/C_\text{pi}(T_\phi-T_\text{t})$, with $C_{\text{pi}}$ being the isobaric heat capacity of ice and $L$ the latent heat of solidification. In order to make sure that the fluid dynamics of the water region is the only influencing factor for the ice evolution, the top temperature, $T_\text{t}$ (correspondingly also the Stefan number), both in experiments and simulations is fixed at a typical value for winter, which we select as $T_\text{t}=-$10$^\circ$C and thus the Stefan number Ste $\approx 20$. The bottom plate temperature, $T_b$ (connected to Rayleigh number to be explained below), is varied in a wide parameter regime, i.e., in experiments $3.8^\circ \text{C} \leq T_b \leq 8^\circ \text{C}$ and in simulations $0.5^\circ \text{C} \leq T_b \leq 15^\circ \text{C}$ (typical water temperature in winter).
We employ laboratory experiments to ensure that the simulations capture all relevant aspects of the physics. The results from the experiments act as the validation for the results from the simulations. On the other hand, simulations can provide more detailed information about the investigated system, and also it is easier to change the values of the control parameters in numerical simulations rather than that of the experiments in the laboratory. So we conduct numerical simulations in a wider and more systematic parameter range than that of the experiments.

An important response to the imposed Ra$_\text{e}$ and Ste is the overall heat flux transported vertically from bottom to top. The dimensionless heat flux is Nusselt number, Nu (more details are shown in {\color{blue} \textit{SI Appendix}, section D}). 

\subsection*{The final average ice position}

\begin{figure*}[!htb]
	\centering
	\includegraphics[width=.8\linewidth]{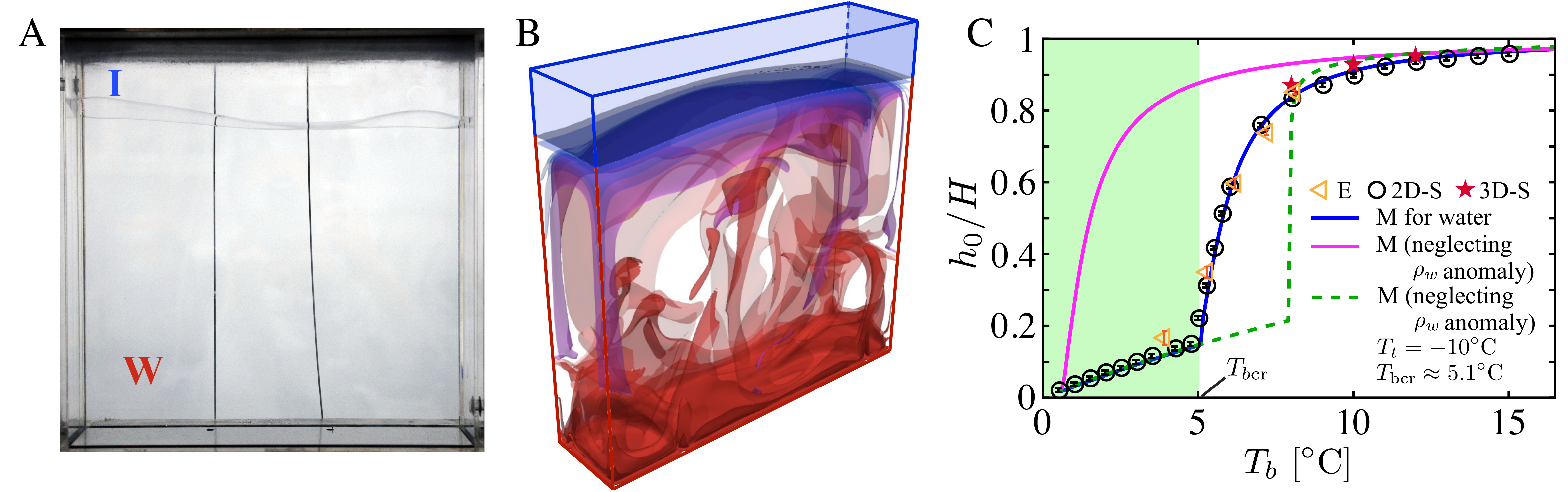}
	\caption{(\textit{A}) Picture of the experimental domain. The system is heated up at the bottom (above the water freezing point $T_\phi = 0^\circ$C) and cooled down from the top (below $T_\phi $). In order to focus on how the fluid dynamics of the water region influences the process of ice formation, the top temperature, $T_\text{t}$, is fixed both in experiments and simulations at a typical value in the winter (which is chosen to be $T_\text{t} = -10^\circ$C). The case shown in (\textit{A}) is at the statistical equilibrium state with $T_\text{b} \approx 8^\circ$C (to give an impression of the approximate timescale of reaching an equilibrium state, it takes a few hours to a few days in the highly convective state and purely conductive state ($T_\text{b}< T_\text{c}$)), ``I'' stands for ice and ``W'' water. (\textit{B}) Visualization of the temperature field across the numerical domain at the statistical equilibrium state (3D-simulation for $T_\text{b} = 8^\circ$C and $T_\text{t} = -10^\circ$C). The blue-colored domain is in the ice region. The ice-water interface is drawn in dark blue.  (\textit{C}) The comparison of spatial-average ice-water interface from the experiments (triangles), the 2D- (circles) and 3D- (stars) simulations, as well as the theoretical model ( legends ``E'', ``2D-S'', ``3D-S'', ``M for water'', ``M (neglecting $\rho_w$ anomaly)'' stand for experiments, two-dimensional simulation, three-dimensional simulation, model for water, model for water without considering the water density anomaly, respectively; blue line: with considering water density anomaly at $T_\text{c}$; violet line: without considering water density anomaly at $T_\text{c}$), and the thermal expansion coefficient $\alpha$ is evaluated at the mean temperature of the investigated range of $T_\text{b}$ ($\sim 7^\circ$C); green dashed line: without considering water density anomaly at $T_\text{c}$, and the thermal expansion coefficient $\alpha$ is a function of temperature, which is evaluated at the mean temperature $T_{mean}$ of the water region for each bottom plate temperature $T_\text{b}$, which is $T_{mean} = (T_\text{b}+T_\phi)/2$. The green-shaded area shows the temperature range corresponding to the diffusive regime of the system. $T_\text{bcr}$ is the critical bottom plate temperature ($T_\text{bcr} \approx 5.1^\circ$C, depending on the model results) above which the system ends up in a convective state. The error bars for the experiments (inside the triangles and comparable to the symbol size) come from the measurement errors (more details are reported in {\color{blue} \textit{SI Appendix}, section B}). The error bars for the simulations (inside the circles) are smaller than the symbol size represent the maximum level of difference between the 2D- and 3D- simulations. }
	\label{FIG1} 
\end{figure*}

We first compare the final average ice position, $h_\text{0}$, which depends on the bottom plate temperature from the experiments, the two dimensional (2D-) and two dimensional (3D-) simulations and from the theoretical model (the details of the model will be discussed later; recall that we use a fixed top temperature ($T_\text{t} = -10^\circ$C) as a typical example, nevertheless, it should be noted that in real natural situations, $h_\text{0}$ may also be influenced by other factors, see also Results and Discussion section). 

Figure~\ref{FIG1}\textit{A} is a photo of the experimental domain at $T_\text{b} \approx 8^\circ$C when the system has reached the statistical equilibrium state. With the same operating conditions, the visualization from the 3D-simulation of the ice position and the temperature field in the fluid phase at the statistical equilibrium state is shown in Fig.~\ref{FIG1}\textit{B}. As shown in Fig.~\ref{FIG1}\textit{A }and \textit{B}, the ice position is similar in the experiment and the numerical simulation in the same condition. Varying bottom plate temperature, $T_\text{b}$, in a large temperature range, the spatially average ice position at the equilibrium state as a function of $T_b$ is shown in Fig.~\ref{FIG1}\textit{C}.  Depending on $T_\text{b}$, the system may end up in a diffusive state (refer to the green shaded area of Fig.~\ref{FIG1}\textit{C}) or in a convective state. There is a good agreement on the height of the spatially average ice-water interface among the experiments, the 2D- and 3D- simulations as well as the theoretical model with considering water density anomaly (see Fig.~\ref{FIG1}\textit{C}, in which ``E'', ``S'', ``M'' stand for experiment, simulation, and model, respectively). However, it is noteworthy that when neglecting the water density anomaly the prediction of ice position from the model (see the violet line and the green dashed line in Fig.~\ref{FIG1}\textit{C}) deviates dramatically from the real value. The violet line is under the assumption that the thermal expansion coefficient, $\alpha$, is a fixed value, which is evaluated at the mean temperature of the investigated range of $T_\text{b}$ ($\sim 7^\circ$C); one may argue that $\alpha$ itself can change with the temperature, then we show the green dashed line, and here $\alpha$ is evaluated at the mean temperature $T_{mean}$ of the water region for each bottom plate temperature $T_\text{b}$, which is $T_{mean} = (T_\text{b}+T_\phi)/2$. Nevertheless, the trend from the model without considering the water density anomaly (the violet line and green dashed line in Fig.~\ref{FIG1}\textit{C}) is very different from the real situation (the blue line in Fig.~\ref{FIG1}\textit{C}).
The key reason is that the stably stratified layer (with temperature ranging from $T_\phi$ to $T_\text{c}$), which results from the density anomaly of water, is crucial for the dynamics of the system. 
The results from the experiments, 2D- and 3D- simulations agree well with each other, which indicates the simulations are reliable, and therefore in the following we will explore the complex nature of the coupled dynamics mostly via 2D- simulations as these allow to more efficiently scrutinize the phenomena in a wide range of parameters.

\subsection*{The coupled dynamics of the ice growth with the fluid motion}
\begin{figure*}[ht]
	\centering
	\includegraphics[width=.9\linewidth]{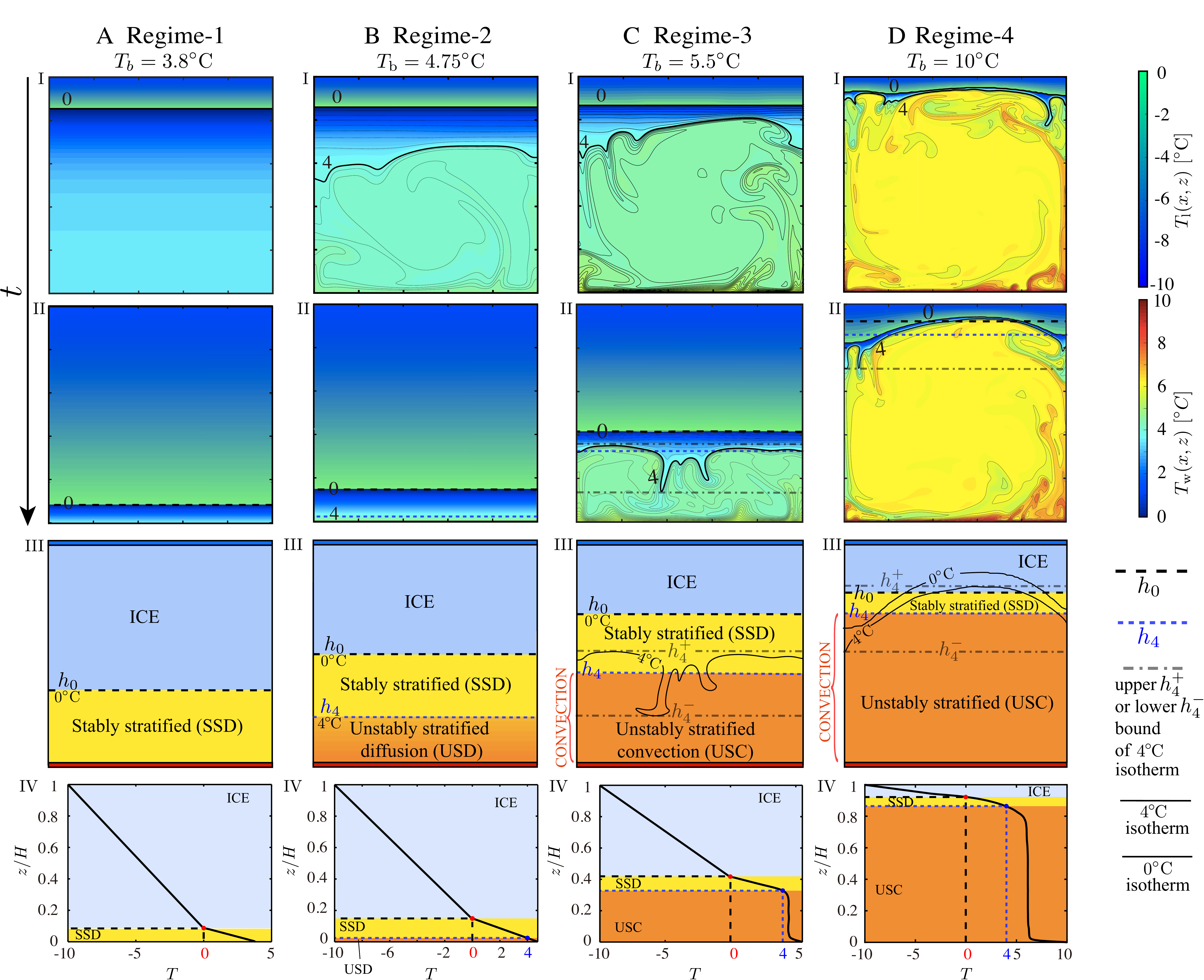}
	\caption{The phenomenology of temporal dynamics and the feature at the statistical equilibrium states in the four regimes. Typical cases visualizations from Regimes-1--4: (\textit{A}) $T_\text{b} = 3.8^\circ$C. (\textit{B}) $T_\text{b} = 4.75^\circ$C (see \textit{SI Movies1}). (\textit{C}) $T_\text{b} = 5.5^\circ$C (see \textit{SI Movies2}). (\textit{D}) $T_\text{b} = 10^\circ$C (see \textit{SI Movies3}). Different time instances of temperature field 
	for four typical regimes of the simulations (panels \textrm{I} and \textrm{II} in (\textit{A})--(\textit{D})).  The sketches on panels \textrm{III} in (\textit{A})--(\textit{D}) depict the coupled different layers of the system at the statistical equilibrium state in Regimes-1--4 respectively, in which the interface (horizontal lines) between neighboring layers (different color shaded areas) are space-average value. The dashed black line is for $h_\text{0}$ (the final average ice position); dotted blue line is for $h_\text{4}$ (the final average $T_\text{c}$-isotherm); thick black curved lines are for instantaneous $0^\circ$C and $T_\text{c}$ isotherms respectively; dash-dotted lines are for the upper bound $h_\text{4}^+$ and lower bound $h_\text{4}^-$ of instantaneous $T_\text{c}$ isotherms. Panels \textrm{IV} in (\textit{A})--(\textit{D}): the temporal and space-average temperature profiles at the statistical equilibrium state corresponding to the four typical cases. In panels \textrm{III}\&\textrm{IV} of (\textit{A})--(\textit{D}), the blue-shaded, yellow-shaded and orange-shaded areas denote ice (ICE), stably stratified layer (SS) and unstably stratified layer (US) respectively. To make the flow structures more visible, two approaches are applied: 1) two colorbars for the temperature field corresponding to ice region($T_\text{I}(x,z)$) and water region ($T_\text{w}(x,z)$) are shown on the right of (\textit{D})\textrm{I}\&\textrm{II}; 2) (\textit{A})\textrm{I}, (\textit{C})\textrm{I}\&\textrm{II}, and (\textit{D})\textrm{I}\&\textrm{II} show more isotherms (thin black lines) except for $0^\circ$C \& $T_\text{c}$ isotherms (thick black lines), which are designed to make the hot and cold plumes more noticeable. }
	\label{FIG2} 
\end{figure*}

To investigate the physical mechanism, we highlight four distinct regimes based on the phenomenology of the equilibrium state as the bottom plate temperature increases from below to above $T_\text{c}$ (Fig.~\ref{FIG2}\textit{A}--\textit{D}). The four regimes that will be considered are as follows, where the first two letters of the acronyms specify the feature of stratification, which can be either the stably stratified (SS) or the unstably stratified (US), and the third letter of the acronyms specifies the mode of heat transport (and fluid motion) which can be either diffusion (D) or convection (C): 1) Regime-1: SSD with flat ice ($T_\text{b}\leq T_\text{c}$) ; 2) Regime-2: SSD + USD with flat ice ($T_\text{c} < T_\text{b}\leq 5.1^\circ \text{C}$); 3) Regime-3: SSD + USC with flat ice ($5.1^\circ \text{C} < T_\text{b} \le 6.9^\circ \text{C}$); and 4) Regime-4: SSD + USC with deformed ice front ($T_\text{b} > 6.9^\circ \text{C}$). 

The boundaries between different regimes depend on the bottom plate temperature (here, around the threshold between each regime, we did simulations with 0.1K increments in order to better identify the transition values). Fig.~\ref{FIG2}\textit{A}--\textit{D} show typical cases from all four regimes from the simulations. 
Next, we discuss the details of the four regimes.

{\textit{\textbf{Regime-1 ($T_b \leq T_\text{c}$)}}}
Figure~\ref{FIG2}\textit{A} shows a typical case in this regime. The system is in a stably stratified state with purely diffusive heat transfer all the way from the beginning (see Fig.~\ref{FIG2}\textit{A}\textrm{I}) till the end (see Fig.~\ref{FIG2}\textit{A}\textrm{II}), the corresponding sketch, which shows different layers at the statistical equilibrium state in the system, can be seen in Fig.~\ref{FIG2}\textit{A}\textrm{III}. The ice-water interface is always flat indicating that the instantaneous 0$^\circ$C isotherm overlaps with the average position of the ice front, $h_\text{0}$. The temperature profiles are linearly dependent on the height both in the ice and water phases, with the different slopes corresponding to the different thermal conductivity in ice and water (Fig.~\ref{FIG2}\textit{A}\textrm{IV}).

{\textit{\textbf{Regime-2 ($T_\text{c} < T_\text{b}\leq 5.1^\circ \text{C}$)}}}
Raising the bottom plate temperature into this regime, the gravitationally unstably stratified layer (from the level of the bottom plate to the spatially average level of $T_\text{c}$ denoted as $h_4$, namely the horizontally average temperature is $T_\text{c}$ at $z=h_\text{4}$, with the temperature ranging from $T_\text{c}$ to $T_\text{b}$) emerges beneath the gravitationally stably stratified layer (from the level of $h_\text{4}$ to $h_\text{0}$ with the temperature ranging from $T_\phi$ to $T_\text{c}$, see the yellow shaded area in  Fig.~\ref{FIG2}\textit{B}\textrm{III}). When $T_\text{b} >T_\text{c}$, in order to know beforehand whether the heat transfer regime is diffusive or convective during the transient state and the statistical equilibrium state, we define the effective Rayleigh number, Ra$_\text{e}$, based on the thickness of the water region from the bottom plate to the spatially average level of $T_\text{c}$ and the corresponding temperature difference, which reads \cite{2020_Bistability}
	\begin{equation}
	\text{Ra}_\text{e}=\frac{(\Delta\rho/\rho_\text{0})g(h_\text{4})^3}{\nu \kappa}=\frac{g\alpha^* (T_\text{b}-T_\text{c})^q (h_\text{4})^3 }{\nu \kappa},
	\end{equation}
with $g$ being the gravitational acceleration, 
$\nu$ the kinematic viscosity, and $\kappa$ the thermal diffusivity. Due to the initial conditions, the system starts from convection in the gravitationally unstably stratified layer, with the $T_\text{c}$ isotherm deformed (see Fig.~\ref{FIG2}\textit{B}\textrm{I}), where Ra$_\text{e} \sim 10^8 \gg \text{Ra}_\text{cr}\approx1708$. ({\color{black}$\text{Ra}_\text{cr}$ is estimated by the linear instability analysis,} which has been intensively validated in the references \cite{pellew1940maintained,dominguez1984marginal,bodenschatz2000recent}). As the ice grows, the effective height, $h_4$, shrinks and Ra$_\text{e}$ consequently decreases. And thus the system ends up at a diffusive state in the entire water layer (SSD+USD) with effective Rayleigh number in US layer Ra$_\text{e} \sim 10$ smaller than Ra$_\text{cr}$. This also explains why the $T_\text{c}$-isotherm becomes flat in the end (see Fig.~\ref{FIG2}\textit{B}\textrm{II}), and the corresponding sketch is shown in Fig.~\ref{FIG2}\textit{B}\textrm{III}. The entire system is in a diffusive state with a linear temperature profile (see Fig.~\ref{FIG2}\textit{B}\textrm{IV}) similar to that in Regime-1.

{\textit{\textbf{Regime-3 ($5.1^\circ \text{C} < T_\text{b}\leq 6.9^\circ \text{C}$)}}}
As $T_\text{b}$ is in Regime-3, with temperature ranging from $5.1^\circ \text{C}$ to $6.9^\circ \text{C}$, there are rich fluid dynamics in the fluid layer below the ice. The system ends up in the convective state with Ra$_\text{e} \sim 10^5$ (see Fig.~\ref{FIG2}\textit{C}\textrm{II}). We can see hot plumes form from the bottom plate. During the lifetime of hot plumes, they detach from the bottom plate shortly after being generated; the plumes accumulate and become coherent plumes, which rise through the bulk region while experiencing heat exchange with the fluid around; if in a classical Rayleigh-B\'enard system they would later on go through the cold boundary layer below the flat $T_\text{c}$-isotherm where they give out most of the energy and slow down to stop, however, in the stably and unstably stratified coupled system presented here, bunches of plumes can impact on and deform the $T_\text{c}$-isotherm because of turbulent bursts.
The $T_\text{c}$-isotherm is no longer flat but develops some spatial variations (see the thick black line in Fig.~\ref{FIG2}\textit{C}\textrm{II}). The region from the spatially average height, $h_4$, of the instantaneous $T_\text{c}$-isotherm to its upper bound, $h_\text{4}^+$, belongs to the gravitationally stably stratified layer but there are also some warmer patches of fluid with the temperature larger than $T_\text{c}$ from the unstably stratified layer. Due to mass conservation, the same amount of fluid, with a temperature smaller than $T_\text{c}$ coming from the gravitationally unstably stratified layer, goes downwards below the level of $h_\text{4}$ (see the downward cold plumes in the region from the lower bound of the instantaneous $T_\text{c}$ isotherm $h_\text{4}^-$ to $h_\text{4}$ in Fig.~\ref{FIG2}\textit{C}\textrm{III}, the flow is convection-dominated in that the estimated Peclet number in the current regime is of order $10^2\sim10^3 \gg 1$; the Peclet number denotes the relative importance of convection with respect to diffusion, and is defined as $Pe = L U / \kappa$\cite{JENKINS2003223}, where $U$ is the characteristic velocity which is taken as the free-fall velocity scale with corrections $U = 0.2\frac{\nu\sqrt{RaPr}}{h_\text{4}}$ for buoyancy driven convection \cite{wang2019self}, and $L$ is the characteristic length scale of the flow which is based on the unstably-stratified layer, and $L$ is the thickness of the unstably-stratified layer, i.e., $L=h_4$). In other words,  due to the non-monotonical behavior of water with respect to the temperature, on average sense there is a stably stratified layer with diffusive heat transfer (SSD, from the level of $h_\text{4}$ to that of $h_\text{0}$) and unstably stratified layer with convective heat transport (USC, from the level of the bottom plate to that of $h_\text{4}$), however instantaneously because of the penetration, there is a strong fluid exchange between SSD and USC as indicated by the deformation of instantaneous $T_\text{c}$-isotherm. Because of the shield of SSD where there is still a horizontal layer with fluid temperature purely smaller than $T_\text{c}$ (from $h_\text{4}^+$ to $h_\text{0}$), the ice-water interface is still flat in Regime-3. In this regime, the temperature profile in the entire water layer is not linear (see Fig.~\ref{FIG2}\textit{C}\textrm{IV}). In the entrainment layer (from the level of $h_\text{4}^-$ to the level of $h_\text{4}^+$) and underneath USC, the temperature profile reflects the turbulence-induced mixing: there is a hot thermal boundary layer attached to the bottom plate, a well-mixed bulk region of nearly uniform temperature, and a cold thermal boundary layer.

{\textit{\textbf{Regime-4 ($T_\text{b} > 6.9 ^\circ \text{C}$)}}}
Upon further increasing $T_\text{b}$ to above $6.9^\circ \text{C}$, the level of upper bound of instantaneous $T_\text{c}$ isotherm, $h_\text{4}^+$, is even higher than the spatially average level of ice position $h_\text{0}$, which indicates strong thinning of the thermal boundary layer by the plumes. On the plume-impact region, ice melts and forms a concave interface due to extra heat input. We can see that there is no horizontally stably stratified layer with fluid temperature purely smaller than $T_\text{c}$ which can shield the ice front from the turbulent convective motion. The $T_\text{c}$-isotherm line is not in a well-defined position, instead, it displays intensive spatial fluctuations due to strong turbulent plumes, resulting in local melting or freezing of the ice front. The water layer consists of a very wide range of USC at the equilibrium state (refer to Fig.~\ref{FIG2}\textit{D}\textrm{II}). The temperature profile is similar to that in Regime-3 but with a much thicker water layer thickness and a much thinner ice layer. {\color{black}The fact of the asymmetrical feature of the thermal boundary layers in the water layer is different from that in the classical Rayleigh-B\'enard convection, which has also been found in \cite{toppaladoddi2018penetrative}, in which they investigated the penetrative convection based on the Prandtl number $Pr = 1$ which is different from the value we used ($\sim 10$). In their work, the reported asymmetrical feature of the thermal boundary layers (fig.~9-11 in \cite{toppaladoddi2018penetrative}) is similar to that found in Regime-3 (Fig.~\ref{FIG2}\textit{C}\textrm{IV}) and Regime-4 (Fig.~\ref{FIG2}\textit{D}\textrm{IV}) of the current study, when the system has the coexistence of the stably and unstably stratified layers.}

In summary, we can see that the heat transfer regimes of diffusion and convection can be even switched dynamically during the evolving process due to the fact that the USC thickness is changing, so the system may end up in a diffusive or convective state depending on the final effective Rayleigh number Ra$_\text{e}$ (which varies with $T_\text{b}$). The statistical equilibrium state depends on the bottom plate temperature, $T_b$. Next, we assess the detailed ice dynamics in a more quantitative perspective. 

\begin{figure}[htbp]
	\includegraphics[width = 0.4\textwidth]{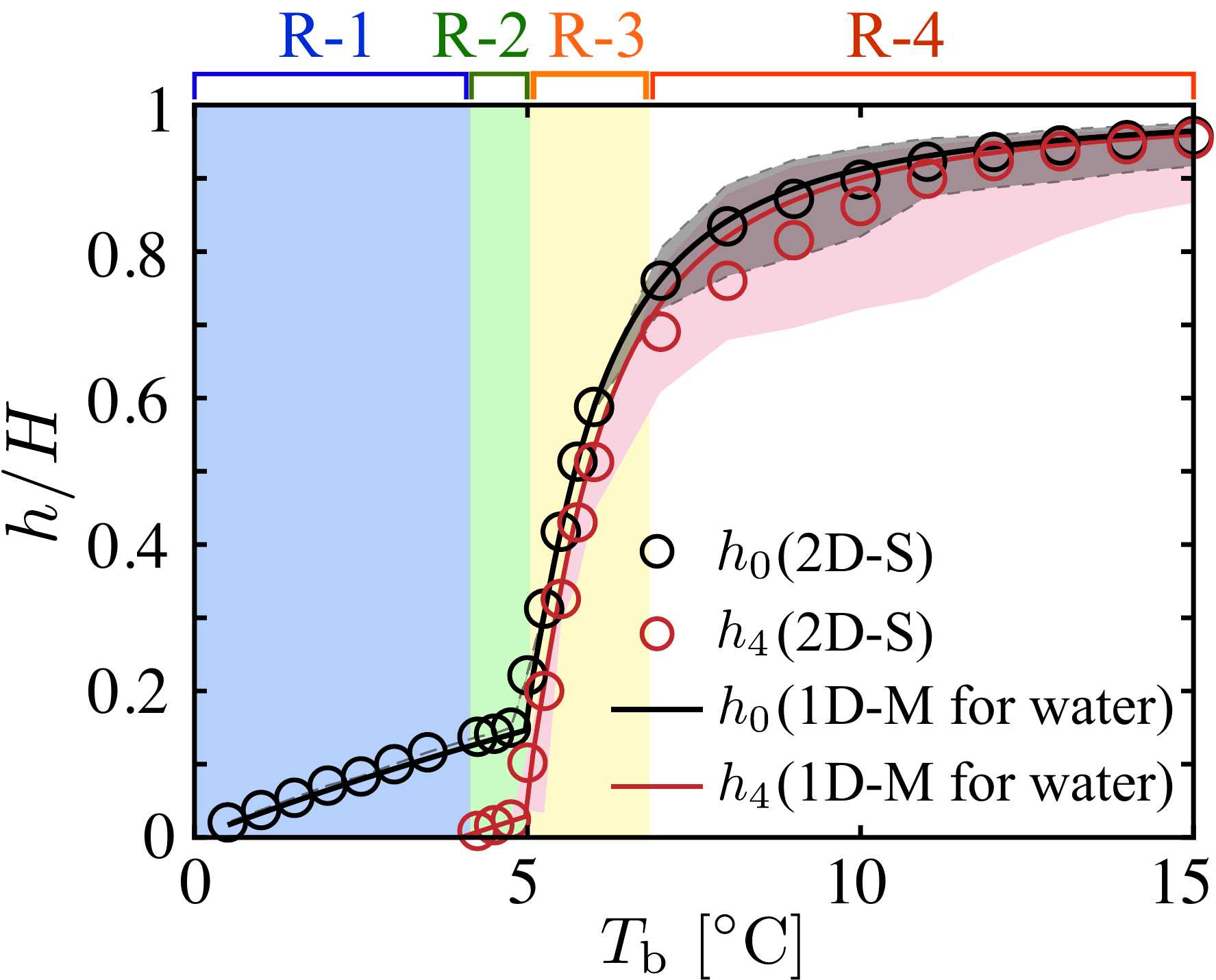}
	\caption{The complex phenomenology emerging from the tight interplay among ice front, stably stratified layer and unstably stratified layer: comparison of the theoretical model (black thick line for $h_\text{0}$, red thick line for $h_\text{4}$) and the simulations featuring the real nature of fluctuations, with circles being the average values (black circles for $h_\text{0}$, red circles for $h_\text{4}$), and black-shaded area and red-shaded area indicating the spatial fluctuation of instantaneous ice-water interface and $T_\text{c}$-isotherm. The blue-shaded area indicates Regime-1 (R-1), green-shaded area Regime-2 (R-2), yellow-shaded area Regime-3 (R-3) and the remaining Regime-4 (R-4). In Regime-4 where $T_\text{b}$ (i.e. Ra$_\text{e}$) is high, the predictions for $h_\text{4}$ deviate a bit from theoretical model due to the intensive interaction among different layers.}
	\label{FIG3} 
\end{figure}

The flow is highly dynamic in the Regime-3 and Regime-4, and the intricate nature of the intensive interaction among the ice front, the entrainment layer, and the unstably-stratified layer leads to high fluctuations of $T_\text{c}$ and $T_\phi$ isotherm varying in a range (see the black-shaded area and the red-shaded area in Fig.~\ref{FIG3}). Nevertheless, the global responses of the system, i.e., the spatial-average thicknesses of the ice-water interface, $h_\text{0}$ (where the horizontally average temperature is 0$^\circ$C), and the, $h_\text{4}$ (where the horizontally average temperature is $T_\text{c}$), match up well to the 1D-model for water (to be discussed below) except for some deviations in the Regime-3 and Regime-4. The ice-water interface and the $T_\text{c}$-isotherm attach to and adjust to each other, which results in a self-organizing large scale circulation, and the overall effects shape the ice front as shown in Fig.~\ref{FIG2}\textit{D}\textrm{II}. 

\subsection*{Theoretical model}

The ice thickness can be properly predicted by taking into account the water density anomaly and the known scaling properties of turbulent thermal convection (namely the Nusselt number-Rayleigh number relation \cite{ahlers2009heat}). Next, we introduce the theoretical model and we consider two situations: 1) for statistical equilibrium states, and 2) for the time-dependent transient states. Here, we assume one dimensional geometry and all the notations are consistent with that in the panels \textrm{III} of Fig.~\ref{FIG2}\textit{A}--\textit{D} (more details about the theoretical model are reported in {\color{blue} \textit{SI Appendix}, section D}).

\textbf{\textit{1) theoretical model for water: statistical equilibrium state}}

When the system has reached the statistical equilibrium state, there is an energy balance between the heat flux through the ice layer and that through the water layer. When $T_\text{b}>T_\text{c}$, the water layer consists of a stably stratified layer (from $T_\phi$ to $T_\text{c}$) and a unstably stratified layer (from $T_\text{c}$ to $T_\text{b}$). Based on the heat flux balance, the average thicknesses of the ice layer ($H - h_\text{0}$), stably stratified layer ($h_\text{0} - h_\text{4}$) and unstably-stratified layer ($h_\text{4}$, exists when $T_\text{b}>T_\text{c}$) at the equilibrium states can be evaluated.

\textit{In the temperature range $T_b \le T_\text{c}$}.
The system is in a diffusive state and independent of the water layer thickness, the total water layer is stably-stratified. According to the conservation of heat flux, we can obtain
\begin{equation}
k_\text{I} \frac{T_\phi - T_t}{H - h_\text{0}} =  k_\text{w}  \frac{T_b -T_\phi}{h_\text{0}}.
\label{eqn_diffusive_lessthan4}
\end{equation}
where $k_\text{I} $ and $k_\text{w}$ are the thermal conductivity of ice and water, respectively. Recall that $T_\phi = 0^\circ$C.
From which we obtain the results on the thicknesses as follows,
\begin{equation}
\left\{
\begin{split}
&H - h_\text{0}=  \frac{-k_\text{I} T_t}{k_\text{w} T_b - k_\text{I} T_t} H,\\
&h_\text{0}= \frac{k_\text{w} T_b}{k_\text{w} T_b - k_\text{I} T_t} H.
\end{split}
\right.
\label{solution_diffusive_lessthan4}
\end{equation}

\textit{In the temperature range $T_b > T_\text{c}$}.
We assume that the interfaces of the ice front and that between the stably stratified and the unstably stratified layers are both flat, based on the heat flux balance between the gravitationally unstably-stratified layer (from the level of the bottom plate to the spatially average level of $T_\text{c}$ denoted as $h_4$, namely the horizontally average temperature is $T_\text{c}$ at $z=h_\text{4}$, with the temperature ranging from $T_\text{c}$ to $T_\text{b}$) and the gravitationally stably-stratified layer (from the level of $h_\text{4}$ to $h_\text{0}$, with the temperature ranging from $T_\phi$ to $T_\text{c}$),
\begin{equation}
\left\{
\begin{split}
&k_\text{I} \frac{T_\phi - T_\text{t}}{H - h_\text{0}} = 
k_\text{w} \frac{T_\text{c}-T_\phi}{h_\text{0}- h_\text{4}},\\
&k_\text{I} \frac{T_\phi - T_\text{t}}{H - h_\text{0}} = 
\text{Nu} k_\text{w}  \frac{T_\text{b}-T_\text{c}}{h_\text{4}}.
\end{split}
\right.
\end{equation}
The model for the heat flux in the unstably stratified layer is in the form of Nusselt number as a function of Rayleigh number (Nusselt number is the dimensionless heat flux defined as $  \text{Nu}=\frac{grad(T)|_{z=0} }{(T_\text{c}-T_\text{b})/h_{4}}$). The Rayleigh number dependence of Nu can be obtained from the simulations and is consistent with that of the classical Rayleigh-B\'enard in the same parameter regime \cite{van_der_poel_stevens_lohse_2013}, suggesting that the model we build is of a general form (more details are reported in {\color{blue} \textit{SI Appendix}, section D}). 
 
By this statistical equilibrium state model, the final ice position, as a function of $T_\text{b}$ (see Fig.~\ref{FIG1}C and Fig.~\ref{FIG3}), and the $T_\text{c}$ isotherm position, as a function of $T_\text{b}$, can be calculated, which show a good agreement with the results from experiments as well as simulations.

\textbf{\textit{2) theoretical model for water: transient state}}

Following the analytical methods for the classical Stefan problem \cite{Alexiades1993Mathematical}, since the time-dependent evolving interface between ice and water (denoted as $z = h_\text{0}(t)$) (where $h_\text{0}(t)$ is the height at which $T_\text{w}(h_\text{0}(t),t) = T_\phi$) is a priori unknown, a part of the solution will be to determine the boundary. As the phase transition occurs, there will be volume change due to the density difference between water and ice as well as the thermal expansion effect. In order to simplify the problem, here we ignore this volume variation. Further, we consider the one-dimension heat transfer problem and assume that the physical properties are invariant with temperature while their values are different for the ice and water phase; the ice-water interface is fixed at phase change temperature $T_\phi $ (recall $T_\phi =0 ^\circ $C). 

When $T_b \le T_\text{c}$, the basic control equations are
\begin{equation}
\frac{\partial T_\text{I}(z,t)}{\partial t} = \alpha_\text{I} \frac{\partial^2 T_\text{I}(z,t)}{\partial z^2}, ~ 0<z<h_\text{0}(t),
\label{ice eqn}
\end{equation}
\begin{equation}
\frac{\partial T_\text{w}(z,t)}{\partial t} = \alpha_\text{w} \frac{\partial^2 T_\text{w}(z,t)}{\partial z^2}, ~ h_\text{0}(t)<z<H,
\label{water eqn}
\end{equation}
where $\alpha$ is the thermal diffusivity, the subscripts ``I'' and ``W'' denote ice and water phase respectively. The boundary conditions read
\begin{equation}
\begin{split}
T_\text{w}(0,t) &= T_\text{b}, \\
\lim_{z \to h_\text{0}(t)^-} T_\text{w}(z,t) &= \lim_{z \to h_\text{0}(t)^+} T_\text{I}(z,t) =T_\phi,\\
T_\text{I}(H,t) &= T_\text{t}.
\end{split}
\label{bc}
\end{equation}
where the superscripts ``$+$'' and ``$-$'' indicate the direction when taking the limit, namely from smaller than $h_\text{0}(t)$ towards $h_\text{0}(t)$ and from larger than $h_\text{0}(t)$ towards $h_\text{0}(t)$, respectively.
The nonlinear energy balance at the ice-water interface is 
\begin{equation}
L \rho_\text{I} \frac{dh_\text{0}(t)}{dt} = k_\text{I} \frac{\partial T_\text{I}(z,t)}{\partial z}|_{z=h_\text{0}(t)^+} - k_\text{w} \frac{\partial T_\text{w}(z,t)}{\partial z}|_{z=h_0(t)^-},
\label{interface eqn}
\end{equation}
From Eqns.~(\ref{ice eqn}, \ref{water eqn}, \ref{bc}, \ref{interface eqn}), we obtain the solutions for temperature distributions in the ice and water,
 \begin{equation}
\begin{split}
&T_\text{w}(z,t) = T_\text{b} - \frac{T_\text{b}}{erfc(\lambda_\text{w})}erfc\bigg(\frac{Z}{2\sqrt{\alpha_\text{w} t}}\bigg),\\
&T_\text{I}(z,t) =T_\text{t}- \frac{T_\text{t}}{erf(\lambda_\text{I})}erf\bigg(\frac{Z}{2\sqrt{\alpha_\text{I} t}}\bigg),
\end{split}
\end{equation}
where $Z = H-z$, erf is the error function ($erfc()x)=1-erf(x)$), and
\begin{equation}
 \lambda_\text{w}=\frac{H - h_0(t)}{2\sqrt{\alpha_\text{w}t}},~ \lambda_\text{I} = \frac{H-h_\text{0}(t)}{2\sqrt{\alpha_\text{I} t}}. 
\end{equation}
 
When $T_b > T_\text{c}$, the effective Rayleigh number can be calculated and the interface energy balance takes the form:

\begin{equation}
L \rho_\text{I} \frac{dh_\text{0}(t)}{dt} = k_\text{I} \frac{\partial T_\text{I}(h_\text{0}(t)^+,t)}{\partial z} + \text{Nu}~k_\text{w} \frac{T_b-T_{\phi}}{h_\text{0}(t)}.
\label{convective interface eqn}
\end{equation}	

Based on Eqn.~(\ref{ice eqn}) and (\ref{convective interface eqn}) with boundary conditions Eqn.~(\ref{bc}), the position of the ice-water interface as a function of time can be solved, and therefore we can predict the temporal evolution of the global icing process of the icing process (see Fig.~\ref{FIG4}B) (more details are reported in {\color{blue} \textit{SI Appendix}, section D}).

\subsection*{Growth dynamics of the ice layer}

\begin{figure}[!ht]
		\centering
	\includegraphics[width = 0.6\textwidth]{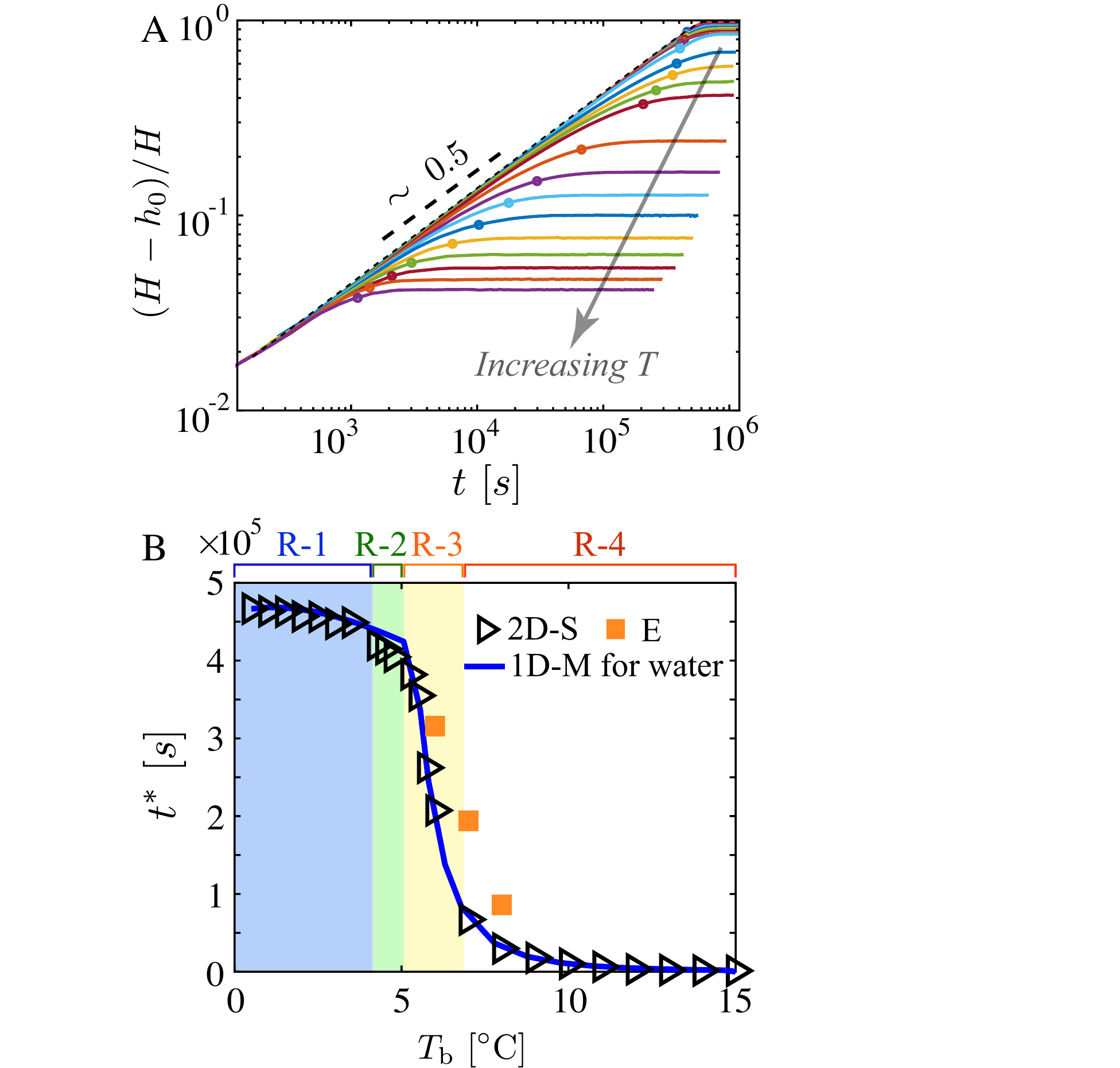}
	\caption{Dynamics of the ice growth: (\textit{A}) Time evolution of the average ice thickness for various $T_\text{b}$. The parameters are $T_\text{t}=-10^\circ \text{C} $ and $0.5^\circ \text{C} \leq T_\text{b} \leq 15^\circ \text{C}$. The gray arrow indicates the direction of increasing $T_\text{b}$. The circles show the saturation time (when ice thickness is increasing to a value of 90\% of that in statistical equilibrium state); (\textit{B}) The saturation time as a function of $T_\text{b}$. The blue-shaded area specifies Regime-1 (R-1), green-shaded area Regime-2 (R-2), yellow-shaded area Regime-3 (R-3) and the remaining Regime-4 (R-4). 
	}
	\label{FIG4} 
\end{figure}

The coupled interactions between the stably and the unstably stratified layers play a major role in determining the final saturation thickness of the ice layer and the time it takes to reach the saturation state (here we define the saturation time, $t^*$, as the time when the ice thickness reaches 90\% of the final statistical equilibrium state thickness). At early stages of the evolution, the conductive heat transfer within the ice layer dominates versus the convective heat transfer within the liquid, because the ice layer is thin where the temperature gradient is large, correspondingly, the conductive heat flux is large. This conduction-dominated early evolution stage yields leading order behaviour of the ice thickness with $(H-h_\text{0})/H \propto t^{0.5}$ at early times even with convective heat fluxes in the liquid (shown in Fig.~\ref{FIG4}\textit{A}). The ice growth deviates from the early leading order behaviour due to the convective flows in the water layer. The saturation time $t^*$ versus the bottom plate temperature is shown in Fig.~\ref{FIG4}\textit{B}, which clearly shows a good agreement between simulations (black symbols) and experiments (orange symbols). 
We also compare the experimental and numerical results with that in the theoretical model. 
They show a good agreement except that there is some deviation in Regimes-2 and 3, which may be due to the complex dynamics around the onset of convection. Further, the coexistence of stably and unstably stratified layers leads to the effective convective region (corresponding to the unstably-stratified layer) smaller than the entire water depth, which may contribute to the discrepancy. Based on the investigated parameter regime, it is revealed that the temperature of the bottom surface has major effects on the icing time.
To give the reader an impression of the physical saturation time scale, for example, when the bottom plate increases from $T_\text{b} = 0.5^\circ$C to $T_\text{b} = 15^\circ$C, the saturation time can vary from a few days to a few hours.

Here, we use a fixed top temperature ($T_\text{t} = -10^\circ$C) as a typical example, nevertheless, it should be noted that in real natural situations, the icing dynamics may also be influenced by the cooling conditions, whole water layer depth and other factors, to which our findings are still applicable and the model is easy to be extended to general situations.

\section*{Conclusions and outlook}

By combing experiments, simulations, and theoretical modeling, we systematically investigated the coupled dynamics between flow and ice growth for different levels of stratification instability (i.e., higher $T_\text{b}$ signifies increasingly unstably stratified). We revealed that the dynamics of the ice thickness can be accurately predicted only by properly taking into account the water density anomaly, in combination with the known GL theory \cite{ahlers2009heat} scaling properties of turbulent thermal convection. We uncovered the rich coupling dynamics among the ice-water interface and the stably- and unstably-stratified layers.

Four regimes were identified depending on $T_\text{b}$ (Regimes-1, 2, 3, and 4), which show the different degrees of interactions with respect to the activities in the water layer.  It is noteworthy that turbulent bursts from the convective unstably stratified layer can penetrate above the $T_\text{c}$ and induce the entrainment layer in Regime-3. 
However, as long as the ice still enjoys the protection of the horizontally continuous stably-stratified layer (where the heat transfers diffusively, SSD), the system terminates with a flat ice-water interface, regardless of the water layer ending up in convection (Regime-3) or conduction state (Regime-1 and 2). Higher thermal intensity (namely high $T_\text{b}$), leads to the deformation of the ice (see Fig.~\ref{FIG2}\textit{D}\textrm{II}), which indicates that some spots of the ice block are thin and vulnerable, and in the case of river or lake ice, these spots may act as initial breakout point. This information is of great importance in de-icing and dredging waterways to provide a more convenient, effective, and smooth freight transportation system in winter.
Further, we showed that, up to a moderate level of turbulence (Regime-1, 2, and 3), the spatially and temporally average ice thickness at the equilibrium state can be well predicted by the theoretical model, suggesting robust predictability of the model with the consideration of the density anomaly.

We found that the ice grows diffusively until the system slowly arrives at the energy balance state and the saturation time can be also well predicted by the theoretical model. Within the investigated parameter regime, the equilibrium time for the ice growth decreases from a few days to a few hours upon increasing $T_\text{b}$, suggesting different environments can tune the final state and its time to consumption. It is noteworthy that our findings can be extended to general situations, such as different environment temperatures and different system sizes, amongst others.

By modifying the thermal condition of the system, the coupling of stably and unstably stratified layers holds promise for regulating local mixing in devices (devoid of moving parts) with respect to contemporary clinical, pharmaceutical, as well as chemical categories, and is ideal for biologically active elements.

The approach followed in this study, which is based on the matching of controlled laboratory-scale experiments with fully resolved direct-numerical simulations sets a standard for future explorations on convection coupled to phase-change problems. We note that the current work has only uncovered a subset of the rich possibilities of ice-water dynamics in terms of the parameter space. In future investigations, we plan to continue by studying the effect of container aspect ratio, ice-water interface inclination, dissolved salt, overburden pressure, the topic of which are of great relevance for better modeling of geophysical and climatological large-scale processes.
 
\section*{Materials and Methods}
Below, we provide basic information on the experiments, theoretical modeling, and numerical simulations performed in this work. Further details and additional figures are provided in {\color{blue} \textit{SI Appendix}}.
\subsection*{Experimental setup}
The turbulent convection coupled with solidification of freshwater experiments were performed in a classical convection setup ({\color{blue} \textit{SI Appendix}, Fig. S1A}). The experimental cell, of rectangular shape, consists of plexiglas sidewalls with height H = $240 mm$ (length $L_x =240$ mm and width $L_y =60$ mm, i.e., aspect ratio $\Gamma = L_x/H = $1.0). The working fluid is confined in between the copper top plate (cooled by circulating bath (PolyScience PP15R-40)) and the copper bottom plate (heated by circulating bath (PolyScience PP15R-40)). In the experiments, ice forms on the top plate and grows in thickness until the system reaches a statistical equilibrium state. During the phase change process, there is a volume change. In order to release the pressure due to volume change induced by phase-change, an expansion vessel is connected to the experimental cell through a tube. The expansion vessel is open to the atmosphere so that the pressure of the experimental cell is kept constant. To avoid evaporation of the water in the expansion vessel, we use a thin layer of silicone oil (immiscible with water) to seal the water surface. By monitoring the water level inside the expansion vessel, the spatial average ice position, $h_\text{0}$ (i.e., the ice thickness is $H-h_\text{0}$), can be calculated for each bottom plate temperature, $T_\text{b}$ {\color{blue} \textit{SI Appendix}, section B}. Six resistance thermistors (44000 series thermistor element, {\color{blue} \textit{SI Appendix}, Fig. S1C}) are embedded into the top and bottom plates, respectively. To control the temperature the setup to avoid the heat exchange between the experimental cell and the environment, there are two kinds of techniques applied: 1) the experimental cell is wrapped in a sandwich structure: insulation foam, aluminum plate, and insulation foam; 2) a PID (Proportional-Integral-Derivative) controller ({\color{blue} \textit{SI Appendix}, Fig. S1B}) is installed to the setup (more details are reported in {\color{blue} \textit{SI Appendix}, section A and section B}). The working fluid is deionized and ultrapure water. Before the experiments, water is boiled twice to degas. Since water density inverses at temperature $T_\text{c}$, here we use the nonmonotonic relationship of density with temperature for water near $T_\text{c}$ and details are reported in {\color{blue} \textit{SI Appendix}, section C}.

\subsection*{Theoretical model}
The theoretical modeling with considering density anomaly for the system is divided into two situations and we assume one-dimensional geometry: 1) for statistical equilibrium states; and 2) for the time-dependent transient states. We also perform theoretical modeling without considering water density anomaly and prove that in this case the ice position can't be predicted properly. More details about the theoretical model are reported in {\color{blue} \textit{SI Appendix}, section D}.

\subsection*{Numerical simulations}
We use Lattice-Boltzmann method (LBM) which is able to capture the turbulent convective dynamics in the water phase and also describe the phase change process at the ice-water interface. {\color{blue} \textit{SI Appendix}, section E} provides more details about the relevant equations that govern phase change, fluid flow, and heat transfer solved by the LBM algorithm.


\paragraph*{Acknowledgments}
	We thank Cheng Wang, Linfeng Jiang, and Varghese Mathai for their help with experiments and insightful discussions. We acknowledge Max Brils for his help in setting up the simulations and for insightful discussions. This work was supported by Natural Science Foundation of China under grant nos 11988102, 91852202, 11861131005 and 11672156, and Tsinghua University Initiative Scientific Research Program under grant no 20193080058. We thank the anonymous reviewers for their constructive comments.

\paragraph*{Author Contributions} CS, EC and FT designed the project; ZW and CS designed the experiments; ZW, FT and EC developed the simulation codes; ZW conducted the experiments; ZW and FT performed the numerical simulations; ZW, EC, CS, FT analyzed and interpreted the data, and wrote the paper. All authors approved the manuscript.



\begin{thebibliography}{10}

\bibitem{meakin2010geological}
Meakin P, Jamtveit B (2010) Geological pattern formation by growth and
  dissolution in aqueous systems.
\newblock {\em Proceedings of The Royal Society A} 466(2115):659--694.

\bibitem{Alboussiere2010Melting}
Alboussiere T, Deguen R, Melzani M (2010) Melting-induced stratification above
  the earth's inner core due to convective translation.
\newblock {\em Nature} 466(7307):744--747.

\bibitem{epstein1983complex}
Epstein M, Cheung F (1983) Complex freezing-melting interfaces in fluid flow.
\newblock {\em Annu. Rev. Fluid Mech.} 15(1):293--319.

\bibitem{worster1997convection}
Worster MG (1997) Convection in mushy layers.
\newblock {\em Annu. Rev. Fluid Mech.} 29(1):91--122.

\bibitem{bilello1968water}
Bilello MA (1968) Water temperatures in a shallow lake during ice formation,
  growth, and decay.
\newblock {\em Water Resources Research} 4(4):749--760.

\bibitem{malm1998bottom}
Malm J (1998) Bottom buoyancy layer in an ice-covered lake.
\newblock {\em Water Resources Research} 34(11):2981--2993.

\bibitem{Veronis1963Penetrative}
Veronis G (1963) Penetrative convection.
\newblock {\em Astrophys. J.} 137:641.

\bibitem{arid2012numerical}
Arid A, Kousksou T, Jegadheeswaran S, Jamil A, Zeraouli Y (2012) Numerical
  simulation of ice melting near the density inversion point under periodic
  thermal boundary conditions.
\newblock {\em Fluid Dynamics \& Materials Processing} 8(3):257--275.

\bibitem{large2014penetrative}
Large E, Andereck C (2014) Penetrative Rayleigh-B{\'e}nard convection in water
  near its maximum density point.
\newblock {\em Physics of Fluids} 26(9):094101.

\bibitem{corcione2015penetrative}
Corcione M, Quintino A (2015) Penetrative convection of water in cavities
  cooled from below.
\newblock {\em Computers \& Fluids} 123:1--9.

\bibitem{ayurzana2016phase}
Ayurzana B, Hosoyamada T (2016) Phase change simulations of water near its
  density inversion point by lattice boltzmann method in {\em Proceedings of
  the 23rd IAHR International Symposium on ice}.
\newblock pp. 1--8.

\bibitem{2019Penetrative}
Wang Q, Zhou Q, Wan Z, Sun DJ (2019) Penetrative turbulent Rayleigh-B{\'e}nard
  convection in two and three dimensions.
\newblock {\em Journal of Fluid Mechanics} 870:718--734.

\bibitem{2020_coupled}
L\'eard P, Favier B, Le~Gal P, Le~Bars M (2020) Coupled convection and internal
  gravity waves excited in water around its density maximum at
  $4^\circ$\text{C}.
\newblock {\em Phys. Rev. Fluids} 5(2):024801.

\bibitem{lecoanet2015numerical}
Lecoanet D, et~al. (2015) Numerical simulations of internal wave generation by
  convection in water.
\newblock {\em Physical Review E} 91(6):063016.

\bibitem{toppaladoddi2018penetrative}
Toppaladoddi S, Wettlaufer JS (2018) Penetrative convection at high Rayleigh-B{\'e}nard.
\newblock {\em Physical Review Fluids} 3(4):043501.

\bibitem{sig94}
Siggia ED (1994) High {{Rayleigh}} number convection.
\newblock {\em Annu. Rev. Fluid Mech.} 26:137--168.

\bibitem{bod00}
Bodenschatz E, Pesch W, Ahlers G (2000) Recent developments in
  Rayleigh-B{\'e}nard convection.
\newblock {\em Ann. Rev. Fluid Mech.} 32(1):709--778.

\bibitem{ahlers2009heat}
Ahlers G, Grossmann S, Lohse D (2009) {Heat transfer and large scale dynamics
  in turbulent Rayleigh-B{\'e}nard convection}.
\newblock {\em Rev. Mod. Phys.} 81(2):503.

\bibitem{lohse2010small}
Lohse D, Xia KQ (2010) {Small-scale properties of turbulent Rayleigh-B{\'e}nard
  convection}.
\newblock {\em Annu. Rev. Fluid Mech.} 42:335--364.

\bibitem{chi12}
Chill\`a F, Schumacher J (2012) New perspectives in turbulent
  {{Rayleigh-B\'enard}} convection.
\newblock {\em Eur. Phys. J. E} 35(7):58.

\bibitem{Madruga2018Dynamic}
Madruga S, Curbelo J (2018) {Dynamic of plumes and scaling during the melting
  of a Phase Change Material heated from below}.
\newblock {\em Int. J. Heat Mass Transf.} 126(B):206--220.

\bibitem{Esfahani2018Basal}
Rabbanipour~Esfahani B, Hirata SC, Berti S, Calzavarini E (2018) Basal melting
  driven by turbulent thermal convection.
\newblock {\em Phys. Rev. Fluids} 3(5):053501.

\bibitem{favier_purseed_duchemin_2019}
Favier B, Purseed J, Duchemin L (2019) {Rayleigh-B\'enard convection with a
  melting boundary}.
\newblock {\em J. Fluid Mech.} 858:437–473.

\bibitem{Satbhai2019Comparison}
Satbhai O, Roy S, Ghosh S, Chakraborty S, Lakkaraju R (2019) {Comparison of the
  quasi-steady-state heat transport in phase-change and classical
  Rayleigh-B\'enard convection for a wide range of Stefan number and Rayleigh
  number}.
\newblock {\em Phys. Fluids} 31(9):096605.

\bibitem{davis1983pattern}
Davis SH, M{\"u}ller U, Dietsche C (1983) {\em Pattern selection in
  single-component systems coupling B{\'e}nard convection and solidification}.
\newblock (Kernforschungszentrum).

\bibitem{dietsche1985influence}
Dietsche C, M{\"u}ller U (1985) Influence of b{\'e}nard convection on
  solid--liquid interfaces.
\newblock {\em Journal of Fluid Mechanics} 161:249--268.

\bibitem{vasil_proctor_2011}
Vasil GM, Proctor MRE (2011) Dynamic bifurcations and pattern formation in
  melting-boundary convection.
\newblock {\em J. Fluid Mech.} 686:77--108.

\bibitem{2020_Bistability}
Purseed J, Favier B, Duchemin L, Hester EW (2020) {Bistability in
  Rayleigh-B\'enard convection with a melting boundary}.
\newblock {\em Phys. Rev. Fluids} 5(2):023501.

\bibitem{Sugawara2000The}
Sugawara M, Irvine TF (2000) The effect of concentration gradient on the
  melting of a horizontal ice plate from above.
\newblock {\em Int. J. Heat and Mass Transf.} 43(9):1591--1601.

\bibitem{Sugawara2007Visual}
Sugawara M, et~al. (2007) Visual observations of flow structure and melting
  front morphology in horizontal ice plate melting from above into a mixture.
\newblock {\em Heat Mass Transf.} 43(10):1009--1018.

\bibitem{mergui2002ice}
Mergui S, Geoffroy S, B\'enard C (2002) Ice block melting into a binary
  solution: coupling of the interfacial equilibrium and the flow structures.
\newblock {\em J. Heat Transfer} 124(6):1147--1157.

\bibitem{Dhaidan2015Melting}
Dhaidan NS, Khodadadi JM (2015) Melting and convection of phase change
  materials in different shape containers: A review.
\newblock {\em Renew. Sustain. Energy Rev.} 43:449--477.

\bibitem{hu2017lattice}
Hu Y, Li D, Shu S, Niu X (2017) Lattice boltzmann simulation for
  three-dimensional natural convection with solid-liquid phase change.
\newblock {\em International Journal of Heat and Mass Transfer} 113:1168--1178.

\bibitem{sugawara2008melting}
Sugawara M, Komatsu Y, Beer H (2008) Melting and freezing around a horizontal
  cylinder placed in a square cavity.
\newblock {\em Heat and mass transfer} 45(1):83.

\bibitem{wintertemp}
(1927-2019) Historical lake erie temperatures, https://www.weather.gov.
\newblock {\em National Weather Service}.

\bibitem{magnuson2000historical}
Magnuson JJ, et~al. (2000) Historical trends in lake and river ice cover in the
  northern hemisphere.
\newblock {\em Science} 289(5485):1743--1746.

\bibitem{magee2016trends}
Magee MR, Wu CH, Robertson DM, Lathrop RC, Hamilton DP (2016) Trends and abrupt
  changes in 104 years of ice cover and water temperature in a dimictic lake in
  response to air temperature, wind speed, and water clarity drivers.
\newblock {\em Hydrology and Earth System Sciences} 20(5):1681--1702.

\bibitem{preston2016climate}
Preston DL, et~al. (2016) Climate regulates alpine lake ice cover phenology and
  aquatic ecosystem structure.
\newblock {\em Geophysical Research Letters} 43(10):5353--5360.

\bibitem{magee2017effects}
Magee MR, Wu CH (2017) Effects of changing climate on ice cover in three
  morphometrically different lakes.
\newblock {\em Hydrological Processes} 31(2):308--323.

\bibitem{succi2001lattice}
Succi S (2001) {\em The Lattice-Boltzmann equation: for fluid dynamics and
  beyond}.
\newblock (Oxford university press).

\bibitem{huber2008lattice}
Huber C, Parmigiani A, Chopard B, Manga M, Bachmann O (2008) Lattice-Boltzmann
  model for melting with natural convection.
\newblock {\em Int. J. Heat \& Fluid Flow} 29(5):1469--1480.

\bibitem{Moritz2019An}
Faden M, König-Haagen A, Brüggemann D (2019) An optimum enthalpy approach for
  melting and solidification with volume change.
\newblock {\em Energies} 12(5):868.

\bibitem{chen2017a}
Chen S, Yan YY, Gong W (2017) A simple Lattice-Boltzmann model for conjugate
  heat transfer research.
\newblock {\em Int. J. Heat and Mass Transf.} 107:862--870.

\bibitem{Gebhart1977A}
Gebhart B, Mollendorf JC (1977) A new density relation for pure and saline
  water.
\newblock {\em Deep Sea Research Part II Topical Studies in Oceanography}
  24(9):831--848.

\bibitem{pellew1940maintained}
Pellew A, Southwell RV (1940) On maintained convective motion in a fluid heated
  from below.
\newblock {\em Proceedings of the Royal Society of London. Series A.
  Mathematical and Physical Sciences} 176(966):312--343.

\bibitem{dominguez1984marginal}
Dominguez-Lerma M, Ahlers G, Cannell DS (1984) Marginal stability curve and
  linear growth rate for rotating couette--taylor flow and Rayleigh-B{\'e}nard
  convection.
\newblock {\em The Physics of fluids} 27(4):856--860.

\bibitem{bodenschatz2000recent}
Bodenschatz E, Pesch W, Ahlers G (2000) Recent developments in
 Rayleigh-B{\'e}nard convection.
\newblock {\em Annual review of fluid mechanics} 32(1):709--778.

\bibitem{JENKINS2003223}
Jenkins W (2003) 6.08 - tracers of ocean mixing in {\em Treatise on
  Geochemistry}, eds.{} Holland HD, Turekian KK.
\newblock (Pergamon, Oxford), pp. 223 -- 246.

\bibitem{wang2019self}
Wang Z, Mathai V, Sun C (2019) Self-sustained biphasic catalytic particle
  turbulence.
\newblock {\em Nat. commun.} 10(1):3333.

\bibitem{van_der_poel_stevens_lohse_2013}
van~der Poel EP, Stevens RJAM, Lohse D (2013) Comparison between two- and
  three-dimensional Rayleigh-B{\'e}nard convection.
\newblock {\em Journal of Fluid Mechanics} 736:177–194.

\bibitem{Alexiades1993Mathematical}
Alexiades V, Solomon AD (1993) Mathematical modeling of melting and freezing
  processes.
\newblock {\em Journal of Solar Energy Engineering} 115(2):121.

\end{thebibliography}

\begin{thebibliography}{1}

\bibitem{Gebhart1977A}
Gebhart B, Mollendorf JC (1977) A new density relation for pure and saline
  water.
\newblock {\em Deep Sea Research Part II Topical Studies in Oceanography}
  24(9):831--848.

\bibitem{Esfahani2018Basal}
Rabbanipour~Esfahani B, Hirata SC, Berti S, Calzavarini E (2018) Basal melting
  driven by turbulent thermal convection.
\newblock {\em Phys. Rev. Fluids} 3(5):053501.

\bibitem{2020_Bistability}
Purseed J, Favier B, Duchemin L, Hester EW (2020) {Bistability in
  Rayleigh-B\'enard convection with a melting boundary}.
\newblock {\em Phys. Rev. Fluids} 5(2):023501.

\bibitem{toppaladoddi2018penetrative}
Toppaladoddi S, Wettlaufer JS (2018) Penetrative convection at high rayleigh
  numbers.
\newblock {\em Physical Review Fluids} 3(4):043501.

\bibitem{van_der_poel_stevens_lohse_2013}
van~der Poel EP, Stevens RJAM, Lohse D (2013) Comparison between two- and
  three-dimensional rayleigh–b\'enard convection.
\newblock {\em Journal of Fluid Mechanics} 736:177–194.

\bibitem{Alexiades1993Mathematical}
Alexiades V, Solomon AD (1993) Mathematical modeling of melting and freezing
  processes.
\newblock {\em Journal of Solar Energy Engineering} 115(2):121.

\bibitem{bodenschatz2000recent}
Bodenschatz E, Pesch W, Ahlers G (2000) Recent developments in
  rayleigh-b{\'e}nard convection.
\newblock {\em Annual review of fluid mechanics} 32(1):709--778.

\bibitem{Moritz2019An}
Faden M, König-Haagen A, Brüggemann D (2019) An optimum enthalpy approach for
  melting and solidification with volume change.
\newblock {\em Energies} 12(5):868.

\bibitem{succi2001lattice}
Succi S (2001) {\em The Lattice-Boltzmann equation: for fluid dynamics and
  beyond}.
\newblock (Oxford university press).

\bibitem{huber2008lattice}
Huber C, Parmigiani A, Chopard B, Manga M, Bachmann O (2008) Lattice-Boltzmann
  model for melting with natural convection.
\newblock {\em Int. J. Heat \& Fluid Flow} 29(5):1469--1480.

\bibitem{chen2017a}
Chen S, Yan YY, Gong W (2017) A simple lattice-boltzmann model for conjugate
  heat transfer research.
\newblock {\em Int. J. Heat and Mass Transf.} 107:862--870.

\end{thebibliography}



\newpage
\begin{widetext}
\section*{SI: Supplementary Information}

\section*{Section A: Experimental setup}

Rayleigh-B\'enard convection experiments coupled with solidification of freshwater are performed in a classical convection setup (see Fig. \ref{exp_setup}). Fig. \ref{exp_setup}\textit{A} reports a sketch of the experimental cell, of rectangular shape, consisting of a plexiglas sidewall with height H = 240 mm (length $L_x =240$ mm and width $L_y =60$ mm, i.e., aspect ratio $\Gamma = L_x/H = $1.0). The working fluid is deionized and ultrapure water. Before conducting any experiments, water is boiled twice to degas. The working fluid is confined in between the copper top plate (cooled by a circulating bath, PolyScience PP15R-40) and the copper bottom plate (heated by a circulating bath, PolyScience PP15R-40). The top and bottom plates and the sidewalls are sealed using silicone O-ring. During the experiments, the top plate temperature, $T_\text{t},$ and bottom plate temperature, $T_\text{b}$, are kept constant, with $T_\text{t} < 0^\circ$C and $T_\text{b} > 0^\circ$C. In the experimental measurements, the temperature fluctuations are less than $\pm0.2$K for the bottom plate and $\pm 0.02$K for the top plate. The fluctuations of the top plate temperature are less than that of the bottom plate, and this is because the ice forms on the top plate and the heat transfer mode is conduction, while on the bottom plate, the turbulent flows of the water region can directly affect the temperature of the bottom plate and induce more fluctuations. In order to focus on how the fluid dynamics of the water region influences the ice formation, the top plate temperature $T_\text{t}$ (and therefore the Stefan number, Ste) is fixed in the experiments at a typical value in winter, which we select as $T_\text{t} = -10^\circ$C (i.e., Ste $\approx 20$). The bottom plate temperature $T_\text{b}$ (i.e., Rayleigh number, Ra) is varied in the temperature range of 3.8$^\circ$C $\le T_\text{b} \le 8^\circ$C. In the experiments, ice forms on the top plate and grows in thickness until the system reaches a statistical equilibrium state. During the phase change process, there is a volume change. In order to release the pressure due to volume change induced by phase-change, an expansion vessel is connected to the experimental cell through a tube. The expansion vessel is open to the atmosphere so that the pressure of the experimental cell is kept constant. To avoid evaporation of the water in the expansion vessel, we use a thin layer of silicone oil (immiscible with water) to seal the water surface. By monitoring the water level inside the expansion vessel, the evolution of the spatial average ice thickness can be calculated. {\color{black} So, when the water level doesn't change in the expansion vessel, it is expected that the system has reached the equilibrium state.} Six resistance thermistors (44000 series thermistor element, see Fig. \ref{exp_setup}\textit{C}) are embedded into the top and bottom plates respectively (refer to the black shaded circles on the top and bottom plates in Fig. \ref{exp_setup}\textit{A} for the positions of the thermistors). The experimental cell is wrapped in a sandwich structure: insulation foam, aluminum plate, and insulation foam. There is a PID (Proportional-Integral-Derivative) controller (see Fig. \ref{exp_setup}\textit{B}) to control the temperature of the setup in order to avoid heat exchange between the experimental setup and the environment. 

We have limited our experimental, simulation, theoretical studies to a constant top undercooling temperature of $T_\text{t} = -10^\circ$C, and we find four typical regimes based on the coupling behaviors. The effects of changing the top undercooling temperature $T_\text{t}$ are qualitatively predictable and are not expected to change the occurrence of the four typical regimes, except that the critical bottom heating temperature for the onset of convection will change. Assume the top undercooling temperature is lower, e.g., $T_\text{t} = -20^\circ$C. At the final equilibrium state, if the ice layer thickness remains the same as that in the case $T_\text{t} = -10^\circ$C, so the temperature gradient in the ice layer is higher, and thus the heat transfer rate in the ice layer is higher than that in the water layer. In order to reach the heat transfer balance state, the ice tends to melt and arrives at a thinner ice thickness, correspondingly, the heat transfer rate in the ice layer decreases and the heat transfer rate in the water layer increases (because the effective Rayleigh number which is defined as $\text{Ra}_\text{e}=\frac{(\Delta\rho/\rho_\text{0})g(h_\text{4})^3}{\nu \kappa}=\frac{g\alpha^* (T_b-4)^q (h_\text{4})^3 }{\nu \kappa}$) increases when the water layer thickness increases). Therefore, the critical bottom heating temperature for the onset of convection decreases.

\begin{figure}[ht]
	\centering
	\includegraphics[width = 1\textwidth]{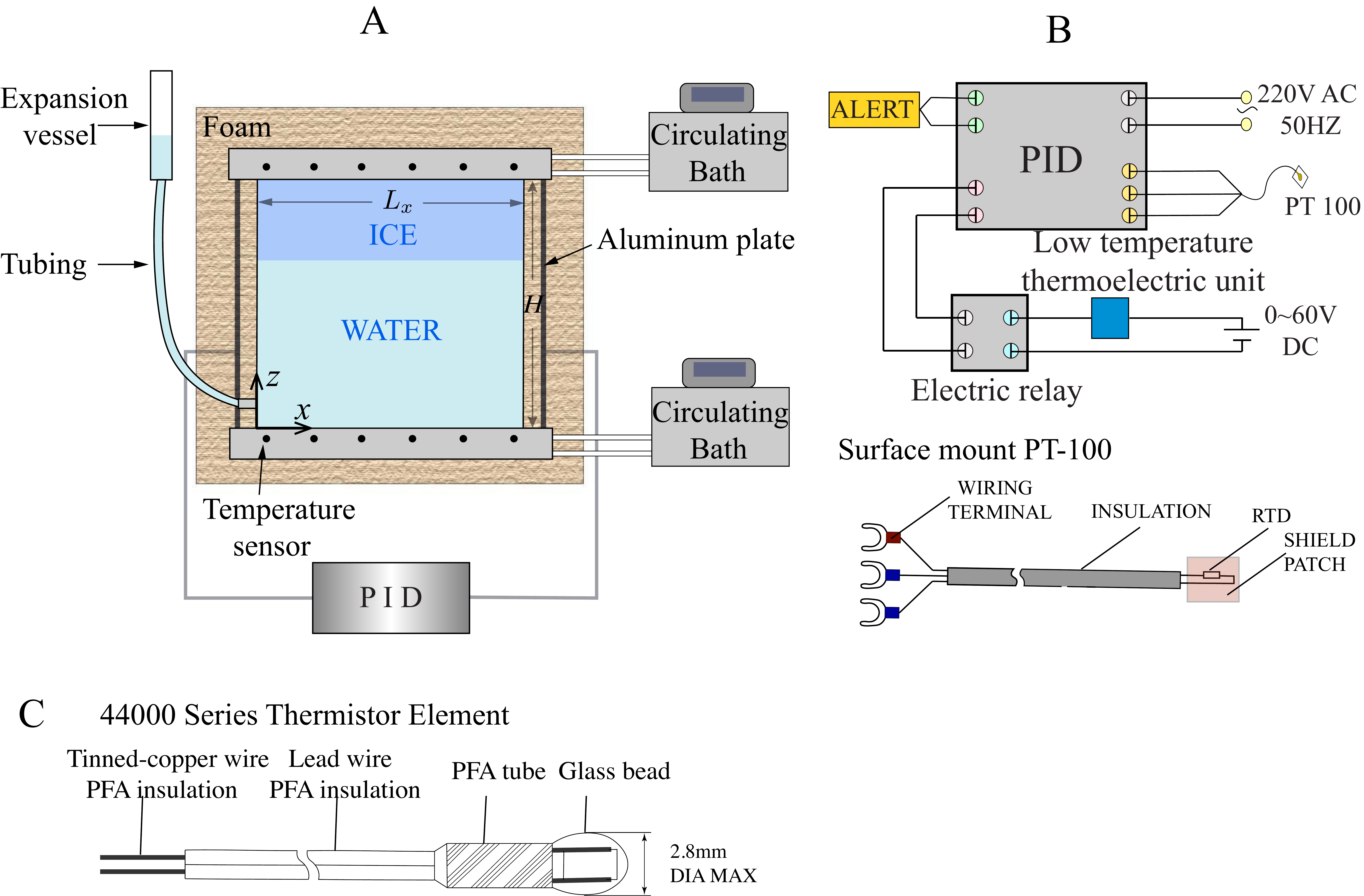}
	\caption{Sketch of the experimental system for Rayleigh-B\'enard convection coupled with solidification of fresh water. (\textit{A}) Experimental cell with insulation facilities. (\textit{B}) The PID (Proportional-Integral-Derivative) controller and the temperature sensor. (\textit{C}) The sketch of resistance thermistor used to monitor the top and bottom plates temperature.}
	\label{exp_setup} 
\end{figure}

\section*{Section B: Calculation of ice thickness as a function of time in experiments}

The evolution of the spatial average ice position, $h_\text{0}$ (the ice thickness is $H-h_\text{0}$), can be calculated for each bottom plate temperature, $T_\text{b}$, by monitoring the water level inside the expansion vessel. 

There are three parts in the system: the Rayleigh-B\'enard convection cell (RB cell, red dashed box), the expansion vessel (EV, green dashed box) and the tube (blue dashed box) which connects the RB cell and the EV (see Fig. \ref{calculation_of_hice}). During the experiments, the mass of water and ice in the system (RB cell + tube + EV) is conserved (the initial mass of the system $m_\text{0}$ is a priori known),  the total volume (the ice volume and the water volume) will change due to isobaric thermal expansion of water and ice formation, both of which will induce a redistribution of the mass in the system. 

The argument of mass conservation yields:

\begin{equation}
\begin{split}
m_\text{0} =\rho_\text{w}(T_\text{m}) \cdot A_\text{RB}\cdot h_\text{0}(t)+ \rho_{\text{I}} \cdot [A_\text{RB}\cdot({\color{black}H}-h_\text{0}(t))] + 
m_{\text{tube}} + 
\rho_\text{w}(T_\text{0}) \cdot V_\text{EV}(t).
\end{split}
\end{equation}

where $\rho_\text{w}$ is the water density as a function of the mean temperature of the water in the RB cell $T_\text{m} = T_\text{b}/2$, $A_\text{RB}$ is the cross sectional area of the RB cell, $\rho_\text{I}$ is the ice density evaluated at the mean temperature of the ice layer $T_\text{t}/2$, $m_\text{tube}$ is the water mass in the connecting tube, and $V_\text{EV}$ is the volume of water in the EV (green dashed box in Fig. \ref{calculation_of_hice}).

So the general form of the ice position, as a function of time $h_\text{0}(t)$, is 

\begin{equation}
\begin{split}
h_\text{0}(t) = \frac{m_\text{0}-\rho_\text{w}(T_\text{0}) \cdot V_\text{EV}(t)-\rho_\text{I}A_\text{RB}{\color{black}H}-m_\text{tube}}{\rho_\text{w}(T_\text{m})A_\text{RB}-\rho_\text{I}A_\text{RB}}
\end{split}
\end{equation}

Next, we estimate the measurement error on the ice position, $h_\text{0}$. The expansion vessel is made of a burette on which there are scales, and therefore the volume of water in the expansion vessel can be read directly. The scale on the expansion vessel has the minimum value of $0.1 ml$ which can lead to accuracy errors when calculating $h_\text{0}$. There is also another factor associated with water evaporation in the expansion vessel that may induce error. It has been mentioned in Section A that the expansion vessel is open to the atmosphere to keep the pressure constant, and we use an oil seal to decrease the evaporation of water from the expansion vessel. To evaluate the evaporation effect, we measure the evaporation rate of water in the expansion vessel on condition of oil seal, which is approximately $1ml$ decrease for three days. On top of these, the minimum and maximum variation in ice position are $0.064 cm$ {\color{black}(for the case when $T_\text{b} = 8^\circ$C )} and $0.28 cm$ {\color{black}(for the case when $T_\text{b} = 3.8^\circ$C)}, which corresponds to 0.5\% {\color{black}(for the case when $T_\text{b} = 8^\circ$C)} to 7\% {\color{black}(for the case when $T_\text{b} = 3.8^\circ$C)} variation of $h_\text{0}$.

\begin{figure}[ht]
	\centering
	\includegraphics[width = 0.6\textwidth]{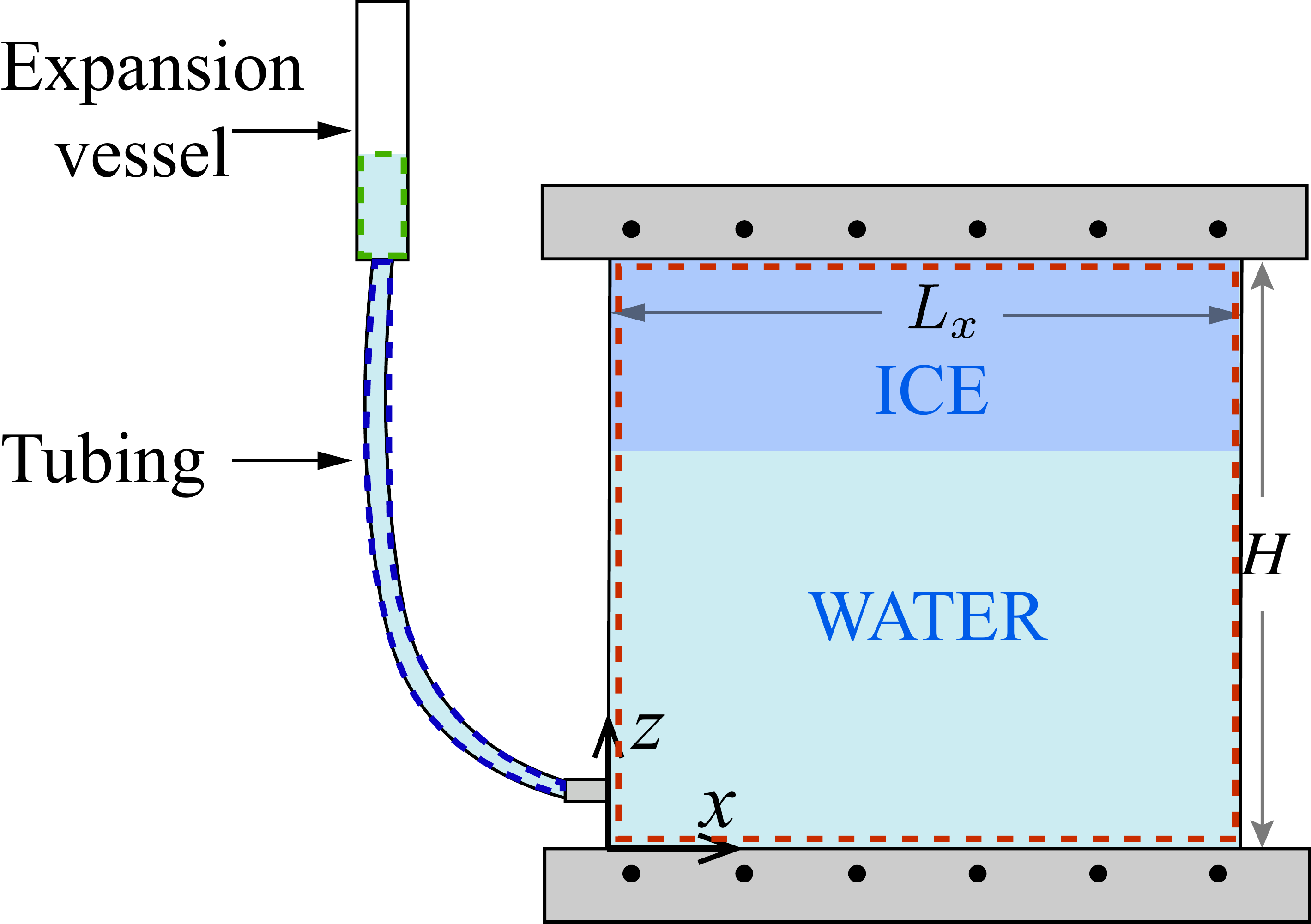}
	\caption{Sketch for the mass conserved region of the system. There are three parts:  Rayleigh-B\'enard convection cell (RB cell, red dashed box), the expansion vessel (EV, green dashed box) and the tube (blue dashed box) which connects the RB cell and EV. }
	\label{calculation_of_hice} 
\end{figure}

\section*{Section C: The nonmonotonic relationship of density with temperature for water near $4^\circ$C}

The working fluid in the experiments is deionized ultrapure water. Since water density inverses at the temperature of $T_\text{c}$ ($\sim 4^\circ$C), here we use the nonmonotonic relationship of density with temperature for water near $T_\text{c}$ from Ref. \cite{Gebhart1977A},
\begin{equation}
\rho_w=\rho_0(1-\alpha^*|T_b-T_\text{c}|^q ),
\label{Ra}
\end{equation}
with $\rho_0 = 999.972 kg/m^3$, $\alpha^* = 9.30\times10^{-6} (K^{-q})$, $q=1.895$.
The density of water, $\rho_w$, as a function of temperature, $T$, is shown in Fig. \ref{rhow}.

\begin{figure}[htb]
	\centering
	\includegraphics[width = 0.5\textwidth]{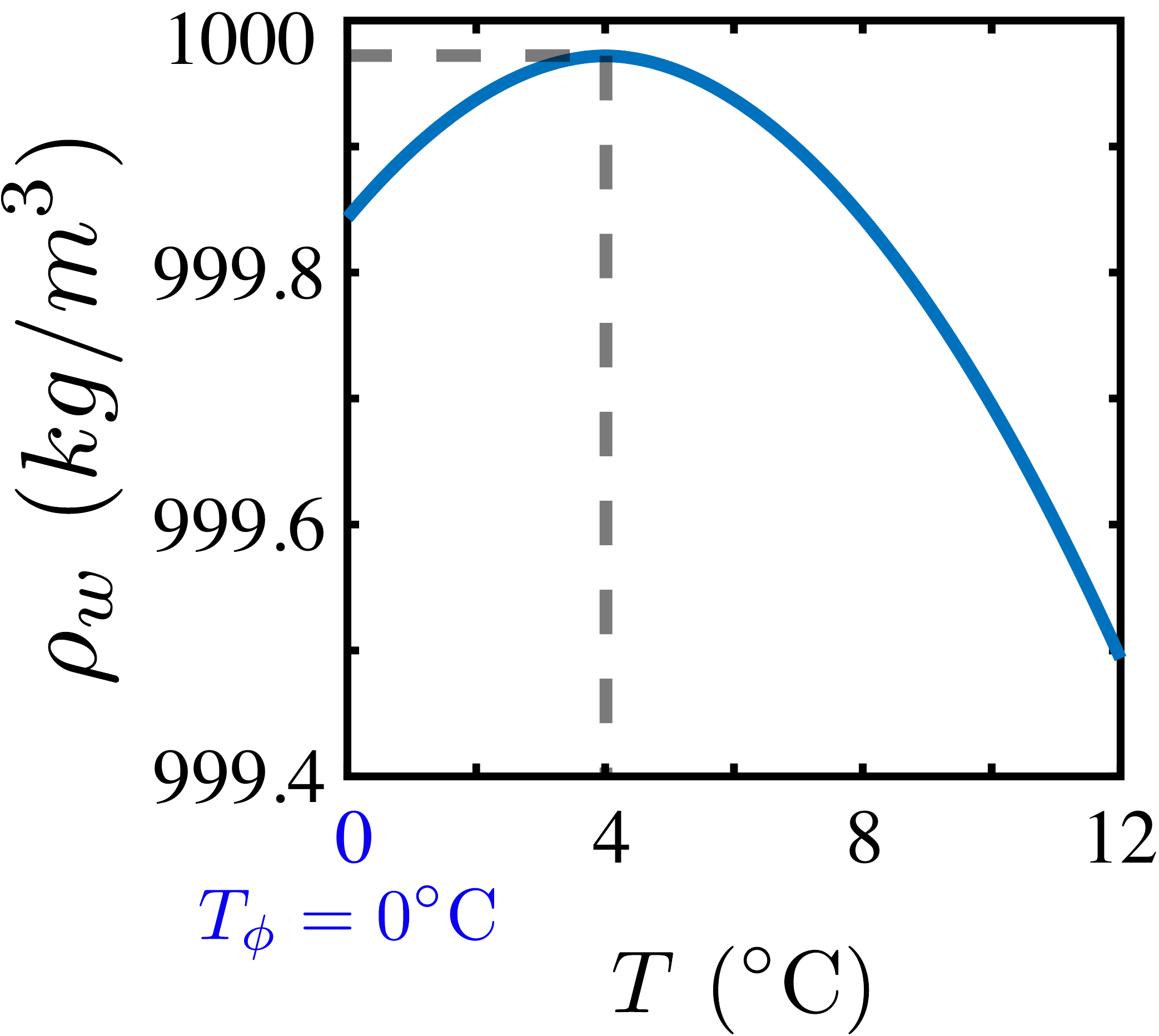}
	\caption{Water density anomaly: the nonmonotonic relationship of density with temperature for cold water near $T_\text{c}$ from Ref. \cite{Gebhart1977A}.}
	\label{rhow} 
\end{figure}

\section*{Section D: Theoretical model}
\subsection*{Theoretical model incorporating water density anomaly}~
In this section, we introduce the theoretical model that accounts for the density anomaly of water. We consider two situations and we assume one-dimensional geometry:

1) for statistical equilibrium states; 

2) for the time-dependent transient states. 

Several relevant previous investigations have also put many efforts into modeling convection. Esfahani et al.\cite{Esfahani2018Basal} employs a linear equation of state for water to model convection with phase change; Purseed et al.\cite{2020_Bistability} numerically studied the Rayleigh-B\'enard convection with phase change (not using water as working fluid) where the melting temperature and the temperature difference between the top and bottom plates are free parameters and are varied; Toppaladoddi \& Wettlaufer. \cite{toppaladoddi2018penetrative} investigated the penetrative convection with a nonlinear equation of state in a system with rigid boundaries.

Next, we discuss the details of the two situations.

\textbf{\textit{1) theoretical model for water: statistical equilibrium state}}

When the system has reached the statistical equilibrium state, there is an energy balance between the heat flux through the ice and that through the water. When $T_\text{b}>T_\text{c}$, the water layer consists of a stably stratified layer (from 0$^\circ$C to $T_\text{c}$) and a unstably stratified layer (from $T_\text{c}$ to $T_\text{b}$).  So there are three kinds of heat flux that balance at the statistical equilibrium state: 1) the diffusive heat flux in the ice layer; 2) the diffusive heat flux in the stably stratified layer; 3) the convective heat flux in the unstably stratified layer, from which we can calculate the average thicknesses of the ice layer ($H - h_\text{0}$), stably stratified layer ($h_\text{0}- h_\text{4}$) and unstably stratified layer ($h_\text{4}$, exists when $T_\text{b}>T_\text{c}$) at the equilibrium state.

An important step is to model the convective heat flux in the unstably stratified layer, which is similar to the classical Rayleigh-B\'enard convection that we will show later. 

Here we define the effective Rayleigh number, Ra$_\text{e}$, based on the unstably stratified layer, which induces the thermal buoyancy driving force when $T_\text{b} >T_\text{c}$, 
\begin{equation}
\text{Ra}_\text{e}=\frac{(\Delta\rho/\rho_\text{0})g(h_\text{4})^3}{\nu \kappa}=\frac{g\alpha^* (T_b-T_\text{c})^q (h_\text{4})^3 }{\nu \kappa}.
\label{Ra}
\end{equation}
where $g$ is the gravitational acceleration, $\nu$ the kinematic viscosity, and $\kappa$ the thermal diffusivity.
Correspondingly, the effective Nusselt number is defined as the heat flux compensated by the diffusive heat flux based on the thickness of the unstably stratified layer $h_\text{4}$ and its temperature difference $(T_\text{b} - T_\text{c}$),
\begin{equation}
\text{Nu}_\text{e}=\frac{grad(T)|_{z=0} }{(T_\text{c}-T_\text{b})/h_{4}}.  
\end{equation}

An empirical fit on the simulation data points similar to Ref. \cite{2020_Bistability} is as follows,

\begin{equation}
\text{Nu}_\text{e} = \left\{
\begin{split}
&1, ~~~~~~~~~~when~ \xi \le 0,\\
&1+C_1\xi, ~when ~1<\xi\le1.23,\\
&C_2 \xi^\beta, ~~~~~when ~\xi>1.23.
\end{split}
\right.
\label{nue}
\end{equation}
with $\xi = (\text{Ra}_\text{e}- \text{Ra}_\text{cr})/\text{Ra}_\text{cr}$, $C_1 = 0.88, C_2=0.27\times \text{Ra}_\text{cr}^\beta$ with $\beta = 0.27$, and all these values are based on the simulation results. 
Nu$_\text{e}$ as a function of Ra$_\text{e}$ is shown in Fig. \ref{nura}, where the simulations results are the red circles and an empirical fit on the simulation data points similar to Ref. \cite{2020_Bistability} given by Eqn.~\ref{nue} is the black line.

\begin{figure}[htb]
	\centering
	\includegraphics[width = 0.6\textwidth]{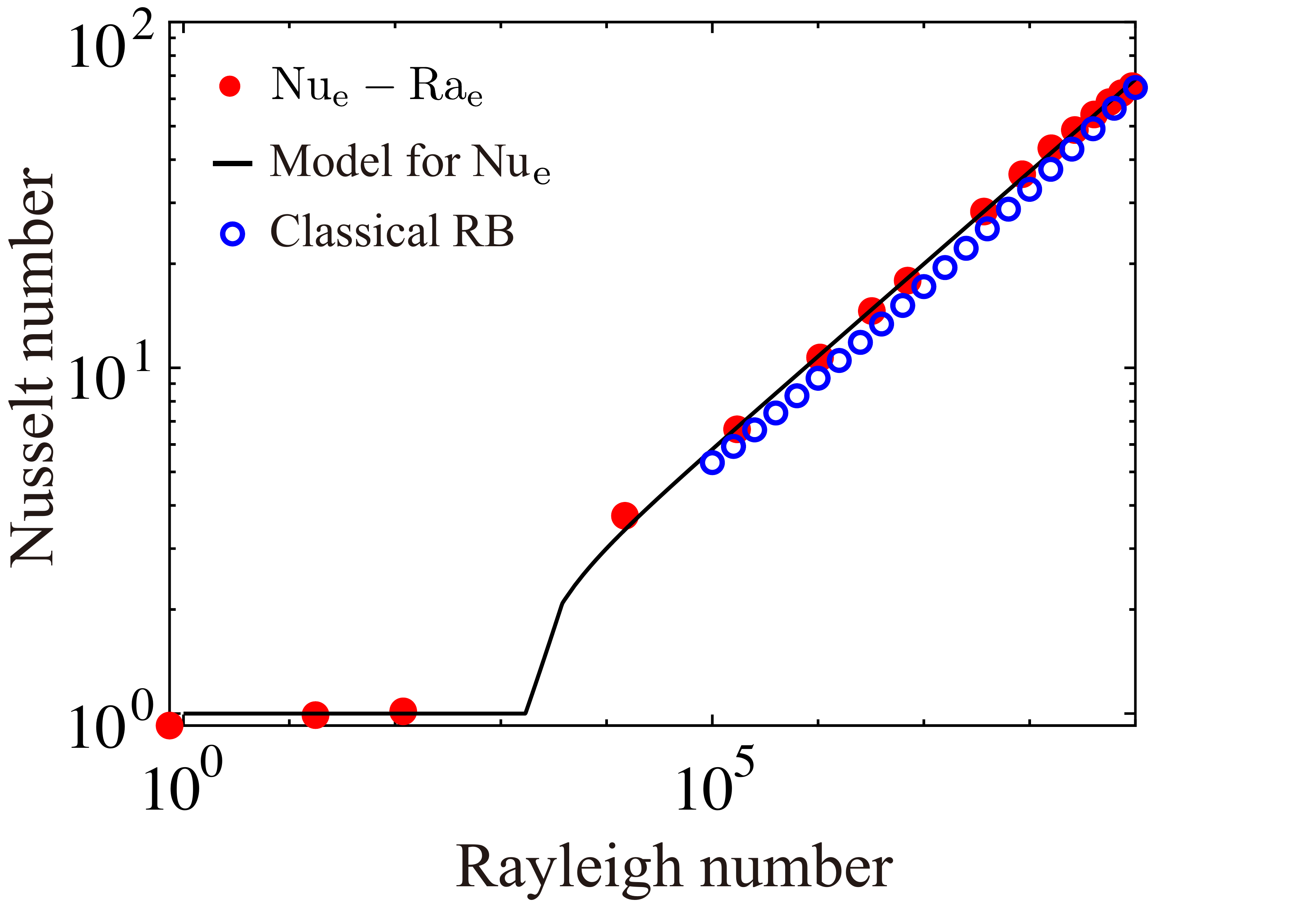}
	\caption{Nusselt number as a function of the Rayleigh number. The red circles are simulation results. The black line is an empirical fit on the simulation data points similar to Ref. \cite{2020_Bistability}. The blue circles are classical Rayleigh-B\'enard convection results from reference \cite{van_der_poel_stevens_lohse_2013}.}
	\label{nura} 
\end{figure}

Fig.~\ref{nura} also reports the numerical results of classical Rayleigh-B\'enard convection from Ref. \cite{van_der_poel_stevens_lohse_2013}. We can see that, despite that our system has different conditions (ice layer, stably stratified layer and unstably stratified layer coexist and couple with one another), there is good agreement on the Nu-Ra relation between the current simulation results and the classical Rayleigh-B\'enard convection, suggesting that the Nu-Ra relation is robust \cite{Esfahani2018Basal}. So we can use the principle in classical Rayleigh-B\'enard convection to model our system.

\textit{In the temperature range $T_b \le T_\text{c}$}.
The system is in a diffusive state and independent of the water layer thickness, the total water layer is stably-stratified,
\begin{equation}
k_{I} \frac{T_\phi - T_t}{H - h_\text{0}} =  k_{w} \frac{T_b -T_\phi}{h_\text{0}}.
\label{eqn_diffusive_lessthan4}
\end{equation}
where {\color{black}$T_\phi $ is the phase change temperature ($T_\phi =0 ^\circ $C),} $k_{I}$ and $k_{w}$ are the thermal conductivity of ice and water, respectively.
From which we obtain the results on the thicknesses as follows,

\begin{equation}
\left\{
\begin{split}
&H - h_\text{0}=  \frac{-k_{I} T_t}{k_{w} T_b - k_{I} T_t} H,\\
&h_\text{0} = \frac{k_{w} T_b}{k_{w} T_b - k_{I} T_t} H.
\end{split}
\right.
\label{solution_diffusive_lessthan4}
\end{equation}

\textit{In the temperature range $T_b > T_\text{c}$}.
We neglect the fact that the interfaces of the ice front and that between the stably and unstably stratified layers can be curved, at the statistical equilibrium state we have
\begin{equation}
\left\{
\begin{split}
&k_{I} \frac{T_\phi - T_\text{t}}{H - h_\text{0}} = 
k_{w} \frac{T_\text{c}-T_\phi}{h_\text{0} - h_\text{4}},\\
&k_{I} \frac{T_\phi - T_\text{t}}{H - h_\text{0}} = 
\text{Nu}_\text{e}k_{w}  \frac{T_\text{b}-T_\text{c}}{h_\text{4}}.
\end{split}
\right.
\end{equation}

\textbf{\textit{2) theoretical model for water: transient state}}

Following the analytical methods of classical Stefan problem \cite{Alexiades1993Mathematical}, since the time-dependent evolving interface between ice and water (denoted as $z = h_\text{0}(t)$, where $h_\text{0}(t)$ is the height at which $T_\text{w}(h_\text{0}(t),t) = 0^\circ$C) is a priori unknown, a part of the solution will be to determine the boundary. As the phase transition occurs, there is a volume change due to the density difference between water and ice as well as the thermal expansion effect. In order to simplify the problem, we ignore here this volume variation. Further, we consider the one-dimension heat transfer problem and assume that the physical properties are invariant with temperature while their values are different for the ice and water phases; the ice-water interface is fixed at the phase change temperature $T_\phi $ (recall $T_\phi =0 ^\circ $C). 

When $T_b \le T_\text{c}$, the basic control equations are
\begin{equation}
\frac{\partial T_\text{w}(z,t)}{\partial t} = \alpha_\text{w} \frac{\partial^2 T_\text{w}(z,t)}{\partial z^2}, ~ 0<z<h_\text{0}(t),
\label{water eqn}
\end{equation}
\begin{equation}
 \frac{\partial T_\text{I}(z,t)}{\partial t} = \alpha_\text{I} \frac{\partial^2 T_\text{I}(z,t)}{\partial z^2}, ~ h_\text{0}(t)<z<H,
\label{ice eqn}
\end{equation}
where $\alpha$ is the thermal diffusivity, the subscripts ``I'' and ``W'' denote ice and water phase respectively. The boundary conditions read
\begin{equation}
\begin{split}
T_\text{w}(0,t) &= T_\text{b}, \\
\lim_{z \to h_\text{0}(t)^-} T_\text{w}(z,t) &= \lim_{z \to h_\text{0}(t)^+} T_\text{I}(z,t) =T_\phi,\\
T_\text{I}(H,t) &= T_\text{t} 
\end{split}
\label{bc}
\end{equation}
where the superscripts ``$+$'' and ``$-$'' indicate the direction when taking the limit, namely from smaller than $h_\text{0}(t)$ towards $h_\text{0}(t)$ and from larger than $h_\text{0}(t)$ towards $h_\text{0}(t)$, respectively.
The energy balance at the ice-water interface is 
\begin{equation}
L \rho_\text{I} \frac{dh_\text{0}(t)}{dt} = k_\text{I} \frac{\partial T_\text{I}(z,t)}{\partial z}|_{z=h_\text{0}(t)^+} - k_\text{w} \frac{\partial T_\text{w}(z,t)}{\partial z}|_{z=h_0(t)^-},
\label{interface eqn}
\end{equation}
where $L $ is the latent heat for solidification of water, $k$ the conductivity.

	Based on the control equations and the corresponding boundary conditions we now derive an explicit expression for the solution.
	We first consider the equation within the water layer Eqn.~(\ref{water eqn}) and  introduce the similarity variable
	\begin{equation}
	\zeta(z,t) = \frac{z}{\sqrt{t}},
	\end{equation}
	and thus the solution is of the form 
	\begin{equation}
	T_\text{w}(z,t) = F(\zeta(z,t)),  
	\label{eqn2_37_ref}
	\end{equation}
	where $F(\zeta(z,t))$ is an unknown function yet to be found.  So the derivatives of $T_\text{w}(z,t) $ are
	\begin{equation}
	\begin{split}
	&\frac{\partial T_\text{w}(z,t)}{\partial t} = \frac{dF}{d\zeta}\frac{\partial \zeta}{\partial t} =  \frac{dF}{d\zeta} \frac{-z}{2t\sqrt{t}},\\
	&\frac{\partial T_\text{w}(z,t)}{\partial z} = \frac{dF}{d\zeta}\frac{\partial \zeta}{\partial z} =  \frac{dF}{d\zeta} \frac{1}{\sqrt{t}},\\
	&\frac{\partial^2 T_\text{w}(z,t)}{\partial z^2} = \frac{1}{\sqrt{t}}\frac{d}{d\zeta}(\frac{dF}{d\zeta})\frac{\partial \zeta}{\partial z}=\frac{1}{t}\frac{d^2F}{d\zeta^2}.
	\end{split}
	\label{derivative}
	\end{equation}
	Substituting Eqn.~(\ref{derivative}) into the heat equation Eqn.~(\ref{water eqn}) gives
	\begin{equation}
	\frac{d^2F}{d\zeta^2} +\frac{\zeta}{2\alpha_\text{w}}\frac{dF}{d\zeta}=0.
	\label{eqn_2_41_ref}
	\end{equation}
	which can be solved with an integrating factor
	\begin{equation}
	M(\zeta) = e^{\int_{h_0(0)}^{\zeta} \frac{h_0(t)}{2\alpha_\text{w}}dh_0}=C_1 e^{\frac{\zeta^2}{4\alpha_\text{w}}}
	\label{eqn_2_42_ref}
	\end{equation}
	where $C_1$ is an integration constant. $M(\zeta)$ in equation (\ref{eqn_2_42_ref}) is multiplied with equation (\ref{eqn_2_41_ref}) we have
	\begin{equation}
	\frac{d^2F}{d\zeta^2}M(\zeta) +\frac{\zeta}{2\alpha_\text{w}}M(\zeta)\frac{dF}{d\zeta}=0,
	\label{prodect_rule}
	\end{equation}
	and by identifying the product rule in equation (\ref{prodect_rule}) we have	
	\begin{equation}
	\frac{d}{d\zeta}\big(M(\zeta)\frac{dF}{\zeta}\big)=0,
	\label{before_inte}
	\end{equation}
	and by integrating equation (\ref{before_inte}) we have 
	\begin{equation}
	M(\zeta)\frac{dF}{d\zeta}=C_2,
	\label{eqn_2_44_ref}
	\end{equation}
	where $C_2$ is an integration constant. The solution of equation (\ref{eqn_2_44_ref}) is 
	\begin{equation}
	F(\zeta) = C\int_{0}^{\zeta}e^{-\frac{h_0^2}{4\alpha_\text{w}}}ds +D
	\label{before error func}
	\end{equation}
	where $D$ is an integration constant. 
	
{\color{black}	From the boundary conditions at $z=0, ~T_\text{w} =T_\text{b}$ and $z=h_0(t), ~T_\text{w} =T_\text{0} $, we can get the temperature distribution within the water which is}
	\begin{equation}
	T_\text{w}(z,t) = T_\text{b} - \frac{T_\text{b}}{erfc(\lambda_\text{w})}erfc\bigg(\frac{Z}{2\sqrt{\alpha_\text{w} t}}\bigg)
	\end{equation}
{\color{black}	with $Z=H-z$, and $ \lambda_\text{w}=\frac{H - h_0(t)}{2\sqrt{\alpha_\text{w}t}}$.	}
	
	With the same method, 
	we can get the temperature distribution within the ice,
	\begin{equation}
	T_\text{I}(z,t) =T_\text{t}- \frac{T_\text{t}}{erf(\lambda_\text{I})}erf\bigg(\frac{Z}{2\sqrt{\alpha_\text{I} t}}\bigg)
	\end{equation}
{\color{black}with $\lambda_\text{I} = \frac{H-h_\text{0}(t)}{2\sqrt{\alpha_\text{I} t}}$. erf is the error function, which is $erf[x] = \frac{2}{\sqrt{\pi}} \int_{0}^{x}e^{-t^2}dx$, and $erfc[x]=1-erf[x]$.}

When $T_b > T_\text{c}$, the water layer consists of stably and unstably stratified layers and the interface between these two layers is $h_\text{4}(t)$, to simplify the problem of estimating the convective heat flux of the water layer, here we define the nominal Rayleigh number Ra and Nusselt number Nu based on the whole water layer from the bottom plate to $h_\text{0}(t)$ with temperature difference $(T_\text{b} - 0^\circ\text{C})$. The definitions for Nu and Ra are as follows:
\begin{equation}
\begin{split}
&\text{Ra}=\frac{g\alpha^* (T_b-0)^q {(h_\text{0}(t))}^3 }{\nu \kappa},\\
& \text{Nu}=\frac{\partial_z T|_{z=0} }{(T_0 - T_b)/h_{0}(t)}. 
\end{split}
\label{nominalRa}
\end{equation}

By comparing the definition of the nominal Rayleigh number, Ra, and Nusselt number, Nu, with the effective Rayleigh number, Ra$_\text{e}$, and effective Nusselt number, Nu$_\text{e}$, we can find the relations in between Ra and Ra$_\text{e}$ as well as Nu and Nu$_\text{e}$ as follows,

\begin{equation}
\begin{split}
&\text{Ra}= \text{Ra}_\text{e} \cdot \varphi_1^q \cdot \varphi_2^3,\\
&\text{Nu}= \text{Nu}_\text{e} \cdot \varphi_1^{-1} \cdot \varphi_2, 
\end{split}
\label{relation}
\end{equation}
with 
\begin{equation}
\begin{split}
& \varphi_1= \frac{T_\text{b} - 0}{T_\text{b} - 4},\\
& \varphi_2 = \frac{h_\text{0}}{h_\text{4}}. 
\end{split}
\label{relation_ratio}
\end{equation}

Using Eqns.~(\ref{nue},\ref{relation}) we can have the model for Nu as a function of Ra so that the convective heat flux based on the whole water layer can be calculated.
The energy balance at the ice-water interface takes the form:
\begin{equation}
L \rho_\text{I} \frac{dh_\text{0}(t)}{dt} = k_\text{I} \frac{\partial T_\text{I}(h_\text{0}(t)^+,t)}{\partial z} + \text{Nu}~k_\text{w} \frac{T_b-T_{\phi}}{h_\text{0}(t)}.
\label{transient_interface}
\end{equation}	

Based on Eqn.~(\ref{ice eqn}) and (\ref{transient_interface}) with boundary conditions Eqn.~(\ref{bc}), the position of the ice-water interface as a function of time can be solved, and therefore we can predict the temporal evolution of the global icing process.

\subsection*{Theoretical model without the water density anomaly}
For the theoretical model without considering the water density anomaly, there is no such thing as nominal or effective Rayleigh number, so here the Rayleigh number Ra$^*$ is defined based on the whole water layer, i.e., the whole water layer thickness $h_\text{0}$ and the corresponding temperature difference $(T_\text{b}-0)$, which is shown as follows,

\begin{equation}
\text{Ra}^*=\frac{g\alpha (T_\text{b}-0) (h_\text{0})^3 }{\nu \kappa},
\label{Ra}
\end{equation}
where $\alpha$ is the thermal expansion coefficient of water evaluated at the mean temperature of the investigated range of $T_\text{b}$ ($\sim 7^\circ$C). 

We can also take the thermal expansion coefficient as a function of temperature, then the Rayleigh number Ra$^*$ reads,
	\begin{equation}
	\text{Ra}^*=\frac{g\alpha|_{at~T_{mean}} (T_\text{b}-0) (h_\text{0})^3 }{\nu \kappa},
	\label{Ra_changingalpha}
	\end{equation}
	where $\alpha|_{at~T_{mean}}$ is evaluated at the mean temperature $T_{mean}$ of the water region for each bottom plate temperature $T_b$, which is $T_{mean} = (T_b+0)/2$.

At the equilibrium state, the energy balances between the diffusive heat flux in the ice layer and the heat flux in the whole water layer, which takes the form:
\begin{equation}
k_{I} \frac{T_\phi - T_\text{t}}{H - h_\text{0}} = 
\text{Nu}_\text{e}k_{w}  \frac{T_\text{b}-T\phi}{h_\text{0}}.
\end{equation}
By using Ra$^*$ to predict the Nusselt number in the Eqn.~\ref{transient_interface} we can solve the equation and get the ice position $h_\text{0}$ for different $T_\text{b}$ just shown in Fig.~1\textit{C} of the main paper.

\section*{Section E: Introduction to the numerical methods: governing equations and numerical simulations}

In this section, we introduce the relevant equations that govern the phase change, the fluid flow, and the heat transfer. 
The governing equations in the water layer are,
\begin{equation}
\begin{split}
&\vec{\nabla} \cdot \vec{u}(x,y,z,t) = 0,\\
& \frac{\partial \vec{u}}{\partial t}  + \vec{u}(x,y,z,t)\cdot \vec{\nabla} \vec{u}(x,y,z,t) = -\frac{\vec{\nabla} p}{\rho_\text{0}} +\nu_w \nabla^2 \vec{u}(x,y,z,t) +\alpha^*g|T(x,y,z,t)-4|^q \textbf{e}_z,\\
&\rho \text{C}_\text{p} \frac{\partial T(x,y,z,t)}{\partial t} +\vec{ \nabla} \cdot (\rho \text{C}_\text{p} \vec{u}(x,y,z,t)T(x,y,z,t)) =\vec{ \nabla} \cdot (k \vec{\nabla} T(x,y,z,t)).
\end{split}
\label{governing_eqn}
\end{equation}
where $\vec{u}(x,y,z,t)$, $p(x,y,z,t)$, $T(x,y,z,t)$ are fluid velocity, pressure, and temperature fields (all temperatures are measured in Celsius), respectively; $\nu_w$, $k$, $\rho$, $\text{C}_\text{p}$, $g$ are the kinematic viscosity of water, the thermal conductivity, the density, the specific heat, and the acceleration of gravity, respectively. When it is in water phase $k = k_{w}$, $\rho = \rho_\text{w} = \rho_0(1-\alpha^*|T_b-4|^q )$, $\text{C}_\text{p}= \text{C}_\text{pW}$, and when it is in ice phase $k = k_{I}$, $\rho = \rho_\text{I}$, $\text{C}_\text{p}= \text{C}_\text{pI}$. All the physical properties of water and ice phase, except for $\rho_w$ are evaluated at the mean temperatures in each phase which are $(T_\text{b}+0)/2$ and $(T_\text{t}+0)/2$, respectively.

The boundary conditions corresponding to the governing equations above are isothermal at the top and bottom plates, no-slip at the bottom plate and at the ice-water interface, adiabatic at the lateral boundaries, and no-slip and freezing (namely, Stefan condition \cite{Alexiades1993Mathematical,bodenschatz2000recent}) at the phase-changing interface. It is noteworthy that in the simulation we use the Boussinesq approximation, which means that the density is regarded as a constant value except for that in the buoyancy term in the momentum equation. Furthermore, we assume the ice and water density remain the same $\rho_I = \rho_w$ to satisfy the incompressibility of the flow. On top of these preconditions, the boundary conditions read:

\begin{equation}
\begin{split}
& T(x,y,0,t) = T_\text{b},\\
& T(x,y,H,t) = T_\text{t},\\
& \textbf{u}(x,y,0,t) = 0,\\
& \textbf{u}(x,y,h_\text{0}(x,y,t),t) = 0,\\
&\frac{\partial {T(0,y,z,t)}}{\partial y} = 0,\\
&\frac{\partial {T(L_x,y,z,t)}}{\partial y} = 0,\\
&{\color{black}L \rho_\text{I} V_n = \textbf{n} \cdot \textbf{q}_\text{w} - \textbf{n}\cdot \textbf{q}_{\text{I}}
}
\end{split}
\label{governing_eqn_bcs}
\end{equation}

where $L$ is the latent heat and $h_0(x,y,t)$ the position vector of a point belonging to the ice-water interface, {\color{black}and $\textbf{q}$ is the heat flux vector, $\textbf{n}$ is a unit normal at the ice-water interface pointing into the liquid. The subscripts I and w refer to the ice and the water, respectively. The heat flux reads $q_\text{I} =  -k_\text{I} \nabla T_\text{I}$ and $q_\text{w} =  -k_\text{w} \nabla T_\text{w}$.}

The boundary condition at the ice-water interface requires particular care due to its time and space dependent character.
So an useful method is to separate the total enthalpy $h$ into sensible heat and latent heat \cite{Moritz2019An}: 

\begin{equation}
\text{h} = \left\{
\begin{split}
&L\phi_w + \text{C}_\text{pI}T, ~~~~~~~~~~~~~~~~~~~~~~~~~~~when~ T<T_\phi,\\
&L\phi_w + \text{C}_\text{pI}T_\phi, ~~~~~~~~~~~~~~~~~~~~~~~~~~when ~T=T_\phi,\\
&L\phi_w + \text{C}_\text{pI}T_\phi + \text{C}_\text{pW}(T-T_\phi), ~~~~~when ~T>T_\phi.
\end{split}
\right.
\label{enthalpy}
\end{equation}
where $T_\phi$ is the phase change temperature ($T_\phi=0$), and $\phi_w(x,y,z,t)$ is the liquid fraction in the system and the relation between $h_0(x,y,t)$  and $\phi_w(x,y,z,t)$ is $h_0(x,y,t) = \int_{0}^{H} \phi_w(x,y,z,t)\, dz$, where $H$ is the height of the investigated domain. In the ice phase, $\phi_w=0$, and in the water phase, $\phi_w=1$, which leads to an additional source term $S_1$ from the latent heat contribution at the ice-water interface in the energy conservation equation of Eqn.~\ref{governing_eqn}. On the other hand, we use the Lattice Boltzmann method (LBM) which is able to capture the turbulent convective dynamics in the water phase and also describe the phase change process at the ice-water interface. The basic principle and formulation has been extensively discussed e.g. in Refs.~\cite{succi2001lattice,huber2008lattice,Esfahani2018Basal}. It is noteworthy that the key to accurately solve such problems is to recover the diffusion term in the energy conservation equation exactly and, similarly to \cite{chen2017a}, we implement the correction when the investigated domain consists of heterogeneous media which lead to another additional source term $S_2$ in the energy conservation equation of Eqn.~\ref{governing_eqn}. So the energy equation with consideration of two source terms $S_1$ and $S_2$ read 

\begin{equation}
\sigma (\rho \text{C}_\text{p})_0 \frac{\partial T(x,y,z,t)}{\partial t} + \vec{ \nabla} \cdot (\sigma (\rho \text{C}_\text{p})_0  T(x,y,z,t)\vec{u}(x,y,z,t)) = \vec{\nabla} \cdot (k \vec{\nabla} T(x,y,z,t)) +S_1+S_2.
\label{source_term}
\end{equation}
 where the first source term is $S_1=- L\rho\frac{\partial \phi_w}{\partial t}$ and the second source term $S_2=-\sigma k \vec{\nabla} T(x,y,z,t) \vec{\nabla} \frac{1}{\sigma} - \frac{\rho \text{C}_\text{p}}{\sigma} T(x,y,z,t) \vec{u}(x,y,z,t) \vec{\nabla} \sigma$. Here $\sigma = \frac{\rho \text{C}_\text{p}}{(\rho \text{C}_\text{p})_0}$ is the ratio of heat capacitance (which is variable and depends on the type of phase, i.e., ice or water) and $(\rho \text{C}_\text{p})_0$ is reference heat capacitance which is taken as constant \cite{chen2017a}.

Next, we explain the benchmarking we have done and provide the resolution used in the simulations.
	
	To sum up firstly, we have done the benchmarking against theoretical solutions as well as simulations of different resolutions. Once the benchmarking is done, we try to shrink the resolution as small as possible, because the ice formation related time scales are on the whole very long. Based on the computation costs and the accuracy of different bottom plate temperatures, we have checked the simulation results of different resolutions to make sure our results are robust. We finally choose the resolution 240 $\times$240, with which 6 $\sim$ 8 lattice nodes can be guaranteed within the thermal boundary layer.
	
	We conducted simulations in the purely conductive regime with different resolutions, i.e., 120$\times$120, 240$\times$240, and 480$\times$480. The ice thickness ratio of different bottom plate temperatures $T_\text{b}$ both from the numerical simulations of different resolutions and from the theoretical modeling is reported in Table S1. 
	
	\begin{table}  
	\renewcommand\arraystretch{1.5}
	\caption{S1 Comparison among different resolutions}  
	\begin{center}
		\begin{tabular*}{10cm}{llllll}  
			\hline  
			$T_b$ ($^\circ$C) &  $h_{ice}^{theory}$&& $h_{ice}^{simulation}$\\
			\hline  
			&&	120$\times$ 120 & 240$\times$ 240 &480$\times$ 480\\  
			\hline  
			0.5  & 0.9830& 0.9833 & 0.9794 & 0.9900\\  
			1.0  & 0.9667& 0.9750 & 0.9625 & 0.9815\\  
			1.5  & 0.9508& 0.9585& 0.9458& 0.9522\\  
			2.0  & 0.9355& 0.9500 & 0.9292 & 0.9396\\  
			2.5  & 0.9206& 0.9417 & 0.9167 & 0.9293\\  
			3.0  & 0.9063& 0.9250 & 0.9000 & 0.9125\\  
			3.5  & 0.8923& 0.9167 & 0.8837 & 0.9002\\  
			\hline  
		\end{tabular*}  
	\end{center}
\end{table}

	The error $\epsilon$ is defined as the relative difference between the results from the theoretical modeling, $h_{ice}^{theory}$ (the theoretical model has been explained in Section D: Theoretical model), and that from the numerical simulations $h_{ice}^{simulation}$
	\begin{equation}
	\epsilon = |\frac{h_{ice}^{theory}-h_{ice}^{simulation}}{h_{ice}^{theory}}| \times 100\% .
	\label{error}
	\end{equation}
	The error $\epsilon$ is shown in Fig. \ref{error_fig}. We can conclude that the simulations with a resolution of 240$\times$240 perform the best with the error of less than 1\% for all cases in the conductive regime. 
	
	\begin{figure}
	\centering
	\includegraphics[width = 0.6\textwidth]{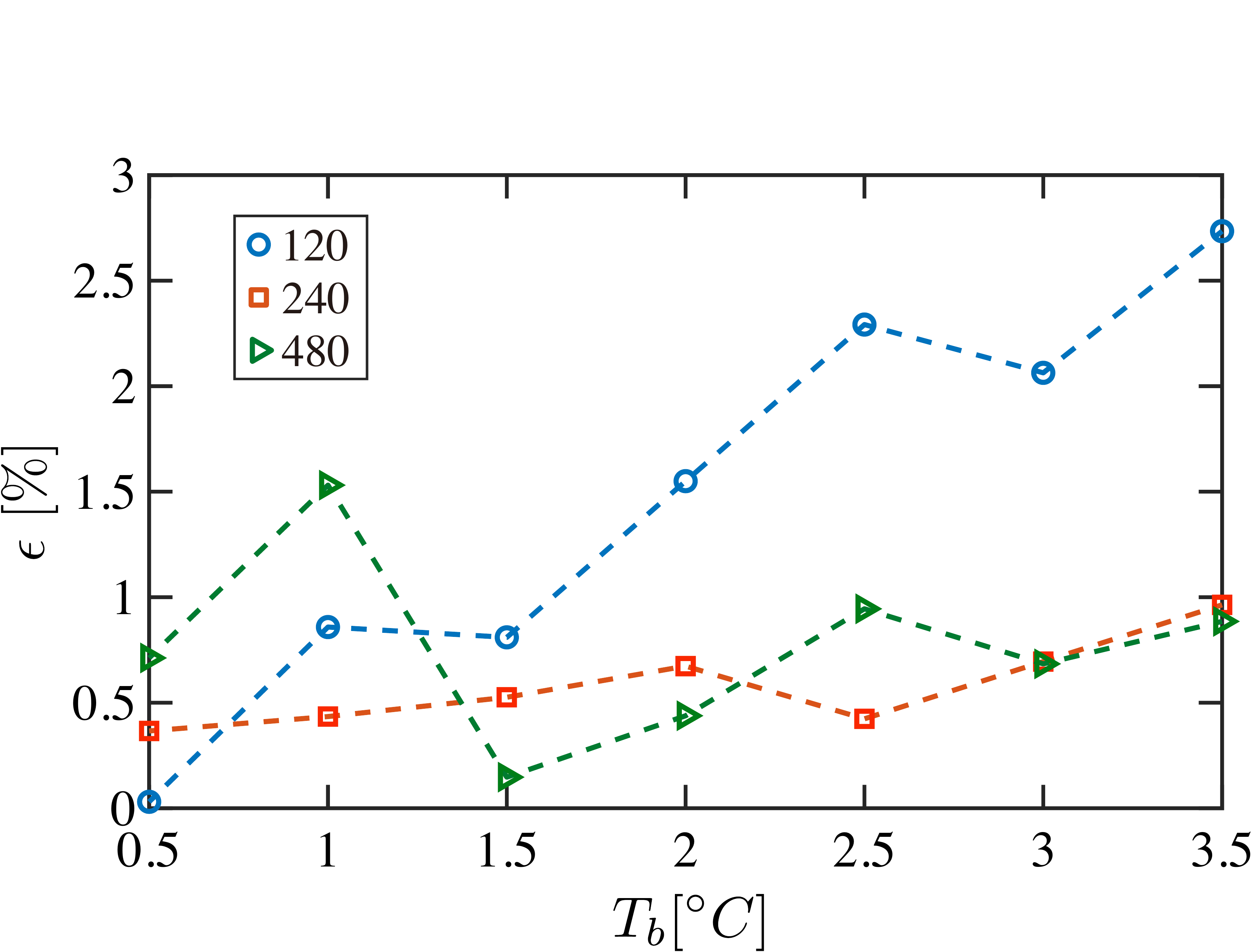}
	\caption{Error of simulation results compared with the theoretical results.  Blue symbols for the resolution 120$\times$120; red symbols for the resolution 240 $\times$240;  green symbols for the resolution 480 $\times$480. Here the dashed lines are drawn as guides to the eye.
	}
	\label{error_fig} 
\end{figure}
	
	It should be noted that when the system is purely conductive, the time it takes to reach an equilibrium is so long that the computation costs are high with increasing resolutions. So it is essential to shrink the size of the system, which is reasonable and luckily possible because in the purely conductive regime the heat transfer behavior is just linear and thus the accuracy can be maintained even with smaller resolutions (for example with a resolution 120 $\times$ 120, the error between the simulation results and the theoretical results is within 3\%). Instead, when the bottom temperature is larger than $4^\circ$C, there is a stably-stratified layer coupling with the unstably-stratified layer, especially in Regime-4, and the flow in the system is highly turbulent. Now the simulation needs bigger systems to maintain the accuracy, but at the same time in this regime the dynamics is much faster. Based on the computation costs and the accuracy, we finally choose the resolution 240 $\times$240, with which 6 $\sim$ 8 lattice nodes can be guaranteed within the thermal boundary layer. So all the simulations reported in the manuscript refer to the resolution 240 $\times$ 240.

\end{widetext}

\end{document}